\documentclass{elsart}
\usepackage{epsfig}
\usepackage{amssymb}
\usepackage[latin1]{inputenc} %specifica la codifica con cui è scritto questo codice sorgente
\usepackage{graphicx} %per inserire le immagini
\usepackage{subfig} %per affiancare le figure
\usepackage{color} %per colorare il testo

\graphicspath{{immagini/}}

\newcommand{\Didasc}{\itshape}

\begin{document}

\begin{frontmatter}
\title{Selective altruism in collective games} 
\author{Dario A. Zappalà}
\ead{dario.a.zappala@gmail.com}
\address{Dipartimento di Fisica e Astronomia - Università di Catania, Via S. Sofia 64, I-95123, Catania, Italy}
\author{Alessandro Pluchino}
\ead{alessandro.pluchino@ct.infn.it}
\address{Dipartimento di Fisica e Astronomia - Università di Catania, and INFN sezione di Catania, Via S. Sofia 64, I-95123, Catania, Italy}
\author{Andrea Rapisarda}
\ead{andrea.rapisarda@ct.infn.it}
\address{Dipartimento di Fisica e Astronomia - Università di Catania, and INFN sezione di Catania, Via S. Sofia 64, I-95123, Catania, Italy}

\begin{abstract}
We study the emergence of altruistic behaviour in  collective games. In particular, we take into account Toral's version of collective Parrondo's paradoxical games, in which the redistribution of capital between agents, who can play different strategies, creates a positive trend of increasing capital. In this framework, we insert two categories of players, altruistic and selfish ones, and see how they interact and how their capital evolves.
More in detail, we analyse the positive effects of altruistic behaviour, but we also point out how selfish players take advantage of that situation.
The general result is that altruistic behaviour is discouraged, because selfish players get richer while altruistic ones get poorer.
We also consider a smarter way of being altruistic, based on reputation, called ``selective altruism'', which prevents selfish players from taking advantage of altruistic ones.
In this new situation it is altruism, and not selfishness, to be encouraged and stabilized.
Finally, we introduce a mechanism of imitation between players and study how it influences the composition of the population of both altruistic and selfish players as a function of time for different initial conditions and network topologies adopted.
\end{abstract}

\end{frontmatter}

\section*{Introduction}
\addcontentsline{toc}{section}{Introduction}
Everyday we observe several examples of altruistic behaviour in the world around us, at different levels of complexity.
Within a single organism, cells coordinate to keep control over their division and avoid the emergence of cancer.
Structures that we observe in the organism (organs, systems) are the consequence of some kind of cooperative behaviour at cellular level.
Coordination and cooperation occur also  within animal societies: for example, worker ants sacrifice their own fecundity and do not reproduce in order  to serve their queen and colony, while, in a pride of lions, adult females nurse not only their own cubs but also those of other females. Humans help each other in many ways: we see small actions of cooperation in every day life, but also  heroic acts like those of the workers of  Fukushima nuclear power plant, who re-entered the contaminated building  
trying  to bring things back under control.
\\
From the point of view of Darwinian evolution, it seems difficult to understand why such behaviour can exist and be so common. In fact, biological evolution is selection and struggle to survive and reproduce. 
Why should one help another individual, risking to lower his own reproductive success for the benefit of someone else?
How cooperative and altruistic behaviour emerge and diffuse in nature?
\\
Kin selection and inclusive fitness are a first possible explanation of these phenomena \cite{haldane,hamilton}, even if restricted to forms of altruism toward close relatives, which are advantageous since they result to increase the individual's genetic contribution to the next generation.
Another possible mechanism able to answer the previous question lies in the key concept of repeated encounters \cite{buchanan}. If two individuals meet once with no chance of meeting again, the best choice for both is obviously to defect, since they have no reason to trust each other and face the risk to be betrayed. But, in a group where individuals often meet one another, the perspective drastically changes. If they help one another in the moments of need, all of them have an advantage. This is the basis of the so called ``tit for tat'' strategy, frequent in human and animal societies \cite{axelrod}.
In this respect, there is also a strong evidence that natural selection operates among groups, as well as among individuals.
It turns out that groups in which cooperating behaviour is  present are favoured over groups of totally selfish individuals \cite{wilson}.
That is possibly why altruistic behaviour emerge and diffuse.
\\
In the past years, altruism has been also analysed with the help of simple mathematical models and simulations, often in the context of game theory.
Among the many studies going in this direction, we may cite the work of Sigmund\cite{sigmund},  Nowak \cite{nowak1}, Gintis \cite{gintis} and Helbing \cite{helbing,helbing2,helbing3}. 
Along this line, in the present paper we focus on a collective version of Parrondo's games \cite{parrondo}.
In particular, we consider here a variant of this model, developed by Toral \cite{toral}, where we introduce altruistic and selfish players.
We investigate  how the action of altruistic players can create a positive condition for the whole community, but we also see how selfish players can take advantage of that situation, creating in the long run a negative condition for the community. In order to investigate how to prevent this bad outcome and explain the emergence of altruism in real situations, we introduce a new and more refined way of being altruistic, which we call \emph{selective altruism}, inspired by the mechanism of indirect reciprocity. Then we explore through extensive numerical  simulations the effects that it produces, also finding the conditions for its diffusion over the entire population.
\\
The paper is organized as follows.
In Section~1 we briefly recap Parrondo's paradox and some collective versions of Parrondo's games, focusing on Toral's one. In Section~2 we introduce the concept of altruism and elaborate a new specific model, called \emph{Altruism-Selfishness (AS) model}, to take it into account. Then, in Section~3, in order to improve this  naive altruistic strategy, we discuss a new model, the \emph{Selective Altruism-Selfishness (SAS) model}, where we introduce a selective altruistic behaviour. In Section~4 we study the effects of imitation among players in a community of fully interacting individuals. Finally, in Section~5, we extend the previous analysis to several communities of players with different network topologies, comparing the simulation results. Conclusions and final considerations close the paper.

\section{Parrondo's paradox in single and collective games}

\begin{figure}[tbp]
\centering
\includegraphics[width=0.7\columnwidth]{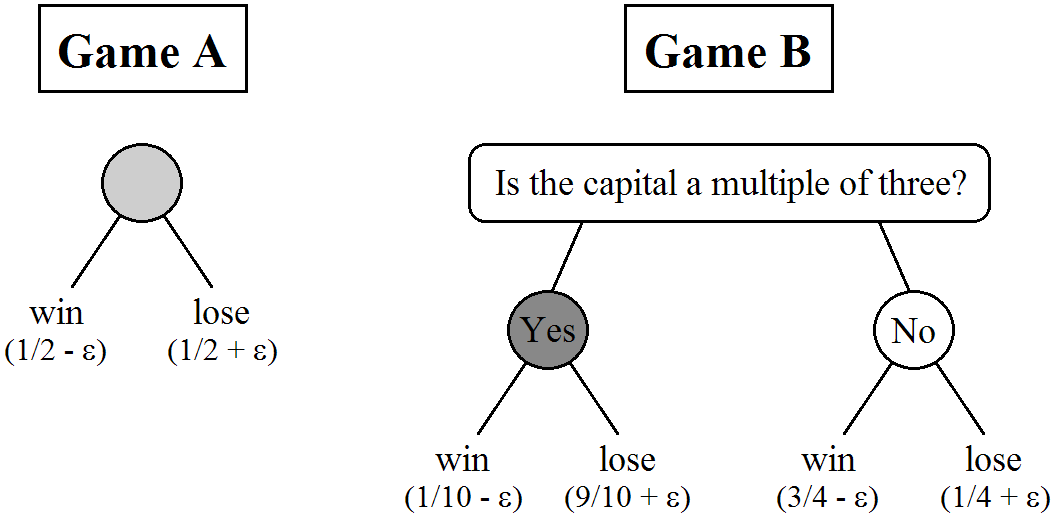}
\caption{{\bf Parrondo's games}. \Didasc Rules of the two original  Parrondo's games A and B, for a single player. Depending on the game and on his capital, the player has to toss a coin with a certain probability of winning and of losing.
The darkness of each coin represents its `badness' for the player.}
\label{Parrondo_rules}
\end{figure}

\begin{figure}[tbp]
\centering
\includegraphics[width=0.7\columnwidth]{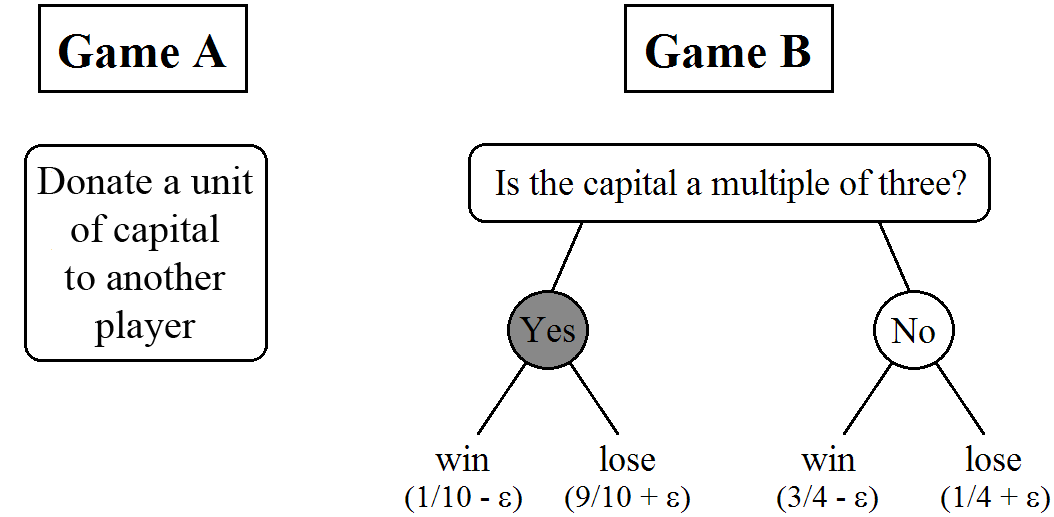}
\caption{{\bf Toral's games}. \Didasc Rules of Toral's games A and B, for a community of $N$ players. Depending on the game and on his capital, the player has to toss a coin with a certain probability of winning and of losing.
The darkness of each coin represents its `badness' for the player.}
\label{Toral_rules}
\end{figure}

Parrondo's paradox \cite{harmer,abbott,parrondo} is a counterintuitive behaviour that takes place in the context of game theory and represents one of the numerous examples of systems where noise and randomness can play a beneficial role \cite{sornette1,sornette2,pluchino1,pluchino2,pluchino3,biondo1,biondo2,biondo3}. 
In particular, Parrondo showed that two losing games can result in a winning trend when played in an alternating or in a random sequence by a single player. 
\\
The games originally described by Parrondo are schematized in Figure~\ref{Parrondo_rules}.
\begin{itemize}
\item \emph{Game~A} consists in a slightly biased coin, with a probability of winning that is less than one half (more precisely $1/2 - \epsilon$). 
\item \emph{Game~B}, on the other hand, consists in two coins, a `bad' one (with a winning probability of $1/10 - \epsilon$) and a `good' one (with a winning probability of $3/4 - \epsilon$): the player tosses the bad coin if his capital is a multiple of three, otherwise he tosses the good coin.
\end{itemize}
As a consequence of game~A or B, at each turn the player can win or lose a unit of capital competing against a casino. It can be proven that the two games, taken singularly, are fair if $\epsilon = 0$ and losing if $\epsilon > 0$. However, a player that alternates (periodically or randomly) the two games (starting from a zero capital) has, on average, a capital that increases with the number of turns, even for small positive values of $\epsilon$. This apparent paradox has been explained by an analogy between the gambling game and a 1-dimensional Brownian motion under the action of a flashing potential \cite{parrondo}. Besides, the paradox can be explained from another point of view: the `profitability' of game~B depends on the probability $\pi_0$ that the capital is a multiple of three. It can be seen that game~A lowers the value of $\pi_0$. In other words, game~A reduces the number of times that the bad coin is used in game~B, making it more profitable.
%@@@@@@@@@@@@@
%@@@@@@@@@@@@@
\\
There are also collective versions of Parrondo's games that produce paradoxical results \cite{dinis,grbehav,parrondo2,greedy,arbtop,netevo}.
More specifically, in refs.~\cite{dinis,grbehav}, the authors show that if every player chooses which game to play aiming only at his own profit, all the community ends up losing capital. On the other hand, in ref.~\cite{parrondo2} it is shown that, if the players choose only for the immediate benefit of the community in the present turn, they still produce a losing trend.
In those cases, it can be seen that chosing randomly between two games or two behavioural patterns avoids falling in the trap of an apparently optimal strategy and produces a winning result.
\\
An interesting collective variant of this scheme, also producing paradoxical results, was introduced by Toral \cite{toral}.
In the capital dependent version of this model (labelled as {\it version I'}  in Toral's paper), depicted in Figure~\ref{Toral_rules}, at each turn just one player $P_i$, in a community of $N$ individuals $\{P_i\}_{i=1,2,\ldots,N}$, is randomly selected for playing and he has to play one of two games.
Game~B is the same as in the original Parrondo's version, whereas game~A consists in the selected player $P_i$ giving away one unit of his capital to another player $P_j$ who is randomly chosen. 
%@@@@@@@@@@@@@
%@@@@@@@@@@@@@
%@@@@@@@@@@@@@
\begin{figure}[tbp]
\centering
\includegraphics[width=\columnwidth]{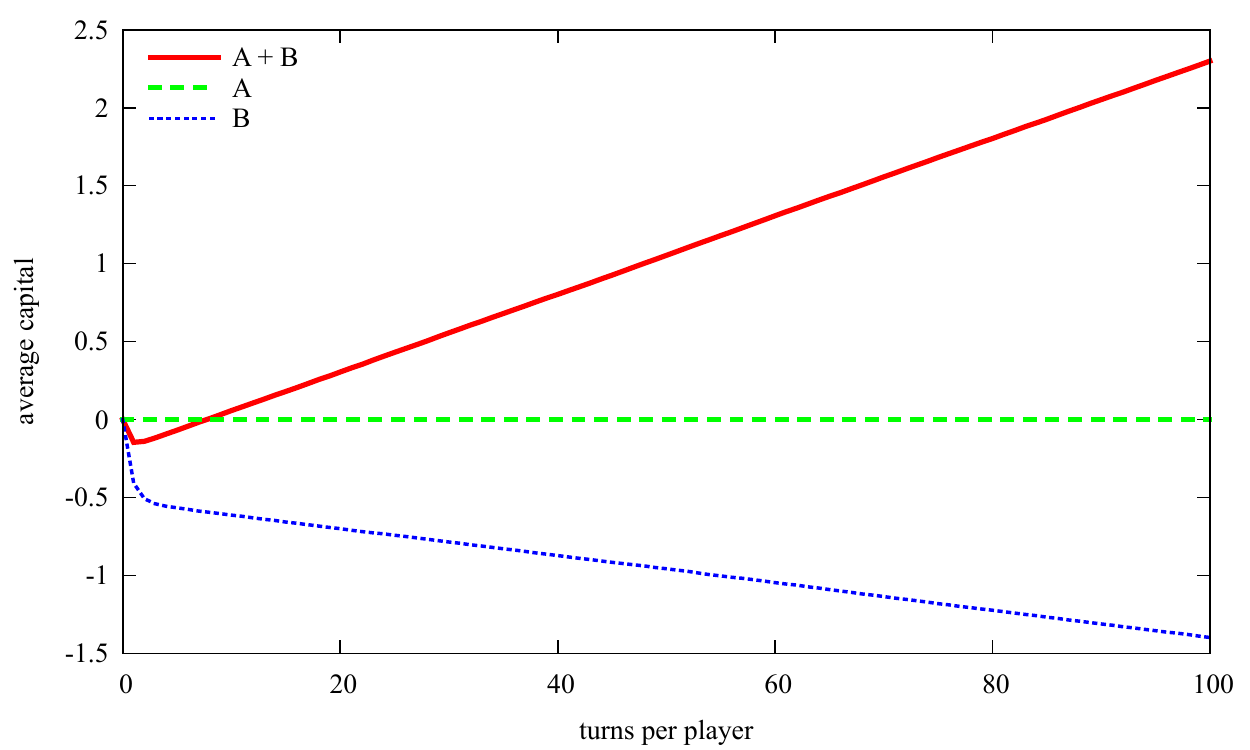}
\caption{{\bf Toral's games}. \Didasc Average capital per player as a function of time (turns per player), in Toral's collective version of Parrondo's games.
We can see how the average capital, starting from zero, evolves in time when all the players play game~A, when they play game~B and when they randomly alternate between the two.
The simulations have been performed with $N = 10,000$ players and $\epsilon = 0.005$, and data have been also averaged over 100 different realizations. See text for further details.}
\label{Toral_orig}
\end{figure}
%@@@@@@@@@@@@@
Game~A does not change collective capital, it just redistributes it among the players. Game~B is losing when played alone, as we already know. The striking result is that, if players randomly alternate between A and B choosing with a probability of $1/2$ for both, they asymptotically increase their capital. This behaviour is evident in Figure~\ref{Toral_orig}, where we can see the results of simulations for a Toral's game with $N=10,000$ fully interacting players (i.e. each player can exchange his capital with anyone else).
The explanation for this paradoxical effect goes along the same lines of the original game: the redistribution of capital lowers the value of $\pi_0$, making game~B more profitable. This means that the redistribution of capital can turn a losing game into a winning game and increase the total capital of the collective group. In this case, ``capital redistribution brings wealth''.
In the next section we will propose a reinterpretation of Toral's collective game in terms of selfish versus altruistic behaviour and we will start to explore the consequences of such assumption. 

%@@@@@@@@@@@@@
\begin{figure}[tbp]
\centering
\includegraphics[width=0.8\columnwidth]{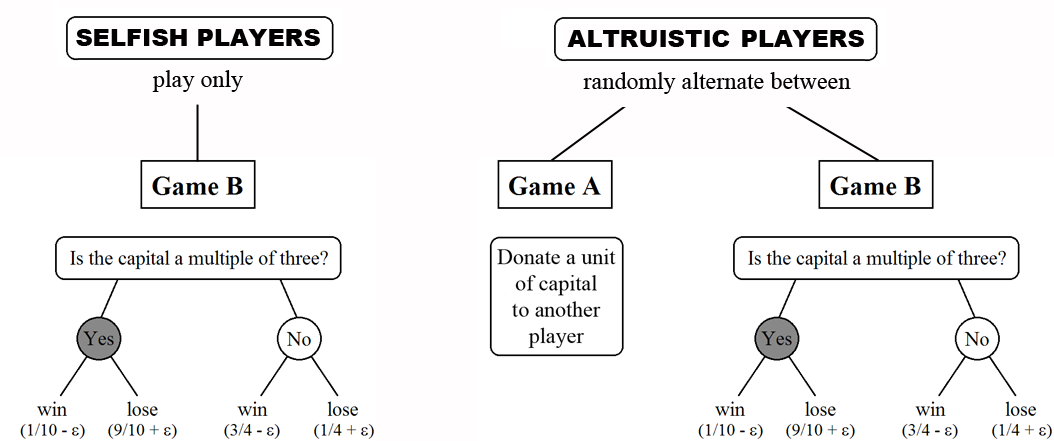}
\caption{{\bf AS model}. \Didasc Rules of our modified version of Toral's collective games for a community of $N$ players. Selfish players play only game B, while altruistic players randomly alternate between game A and game B. }
\label{Altruistic_rules}
\end{figure}
%@@@@@@@@@@@@@

\section{Altruism and its effects: the AS model}

Altruistic behaviour is a topic largely debated in the scientific literature. 
In particular, there is significant interest in studies dealing with the emergence of cooperative behaviour, in which the single agent sacrifices his personal gain for the gain of all the others \cite{nowak1,helbing,helbing2}.
To introduce the concept of altruism in this game context, let us look at Toral's model from another point of view.
\\
It is clear that, from the perspective of a single player, game~A is invariably a losing game: the player loses a unit of capital, donating it to another player.
On the other hand, game~B has some chances of winning: the player competes against a casino and can win or lose a unit of capital.
So, if a player had to choose between the two games caring only for himself, he would have no reason to choose A.
In this respect, we may consider the case in which players play only game~B as a selfish behaviour: no one wants to lose capital for the benefit of the community.
On the other hand, we may regard the case in which players randomly alternate between A and B as an altruistic behaviour: players still play against a casino to increase their capital, but from time to time they sacrifice their personal wealth to favour the game of other players.
\\
Following these interpretations, in this section we will introduce and study a modified version of Toral's model in order to test our intuitions and answer some questions about the stability of altruistic behaviour.
We will refer to this model as \emph{Altruism-Selfishness (AS) model} and we schematically illustrate it in Figure~\ref{Altruistic_rules}.
Let us consider a community of $N$ fully interacting players taking part in the game at the same time and distributed into two classes: 
\begin{itemize}
\item \emph{altruistic players}, indicated as $\{A_i\}_{i=1,2,\ldots,N_{\mathrm a}}$, who alternate randomly between game~A and game~B, so that they can choose from time to time the losing option only to help another player;
\item \emph{selfish players}, indicated as $\{S_i\}_{i=1,2,\ldots,N_{\mathrm s}}$, who play only game~B, because they do not want to give away a part of their capital.
\end{itemize}
Of course, $N_{\mathrm a} + N_{\mathrm s} = N$. 
Each player has an initial capital $C_i(0)=0$ $(i=1,2,\ldots,N)$ that will change in time depending on the played game.
Time $t=0,1,2,\ldots,T$ is a discrete variable indicating the number of played turns (during each turn, all the players play their games in a random order). 

\subsection{Impact of a fraction of altruistic players}

\begin{figure}[tbp]
\centering
\includegraphics[width=\columnwidth]{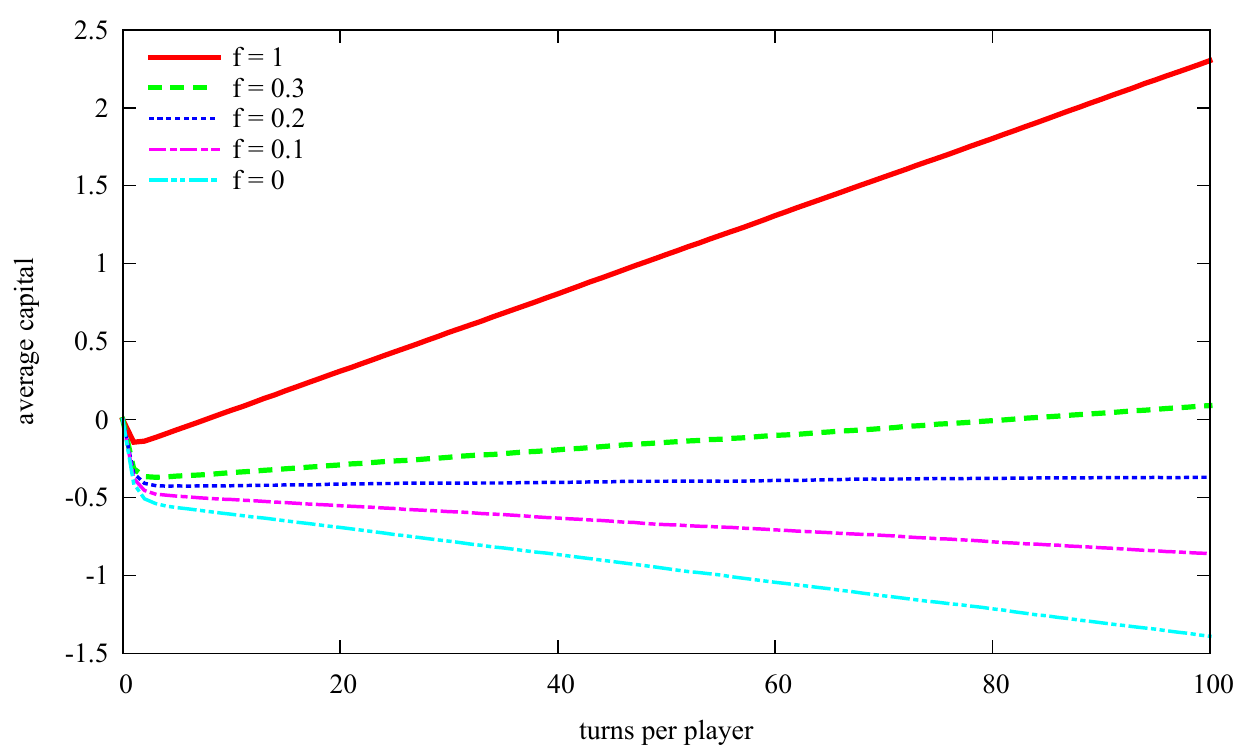}
\caption{{\bf AS model}. \Didasc Time evolution of the average capital $\bar{C}(t)$, for different values (indicated in the legend) of the fraction $f$ of altruistic players. Here we have $N = 10,000$ players and $T=100$ turns per player. The parameter of Parrondo's games is $\epsilon = 0.005$. The results have been also averaged over 50 realizations.}
\label{fraction-capital_evolution}
\end{figure}

Using this newly built model, the first point we want to investigate is what happens when only some of the players are altruistic.
Is it necessary that all the players are altruistic in order to have a winning trend for the collective group?
We are interested in calculating how the average capital $\bar{C}(t)=\frac{1}{N} \sum_{i=1}^N C_i(t)$ evolves in time as a function of the (fixed) fraction $f=N_{\mathrm a}/N$ of altruistic players present in the community.
In Figure~\ref{fraction-capital_evolution} we show the results of a first set of simulations for a community with $N=10,000$ players and for several increasing values of the fraction $f$. 
%@@@@@@@@@@@@@
%@@@@@@@@@@@@@
The case $f = 0$ (i.e. only selfish players, $N_{\mathrm s} = N$) is equivalent to playing only game~B in Toral's model, therefore we observe the expected linear decrease of the average capital $\bar{C}(t)$ as a function of time. On the other hand, the case $f = 1$ (i.e. only altruistic players, $N_{\mathrm a} = N$) corresponds to the random alternation of A and B for all the players in Toral's model, thus it produces, after a short decreasing transient, a linearly increasing average capital. For intermediate values of $f$, we see that the slopes of the straight lines increase almost continuously when the value of $f$ grows. The value $f \sim 0.2$ behaves like a threshold: below that value, the slope is negative and the average capital is decreasing (the game is losing for the collective group); above that value, the slope becomes positive and the average capital starts to increase (the game is collectively winning).
Summarizing, even a small fraction of altruistic players seems sufficient to influence all the community with its positive effects, making the game winning for the collective group and producing an increasing collective capital.
 In other words, if just a few players  donate part of their own capital,  one observes  a movement of wealth within the community sufficient to lower the value of $\pi_0$ and make game~B winning.

\subsection{Disadvantages of being altruistic}

\begin{figure}[tbp]
\centering
\includegraphics[width=\columnwidth]{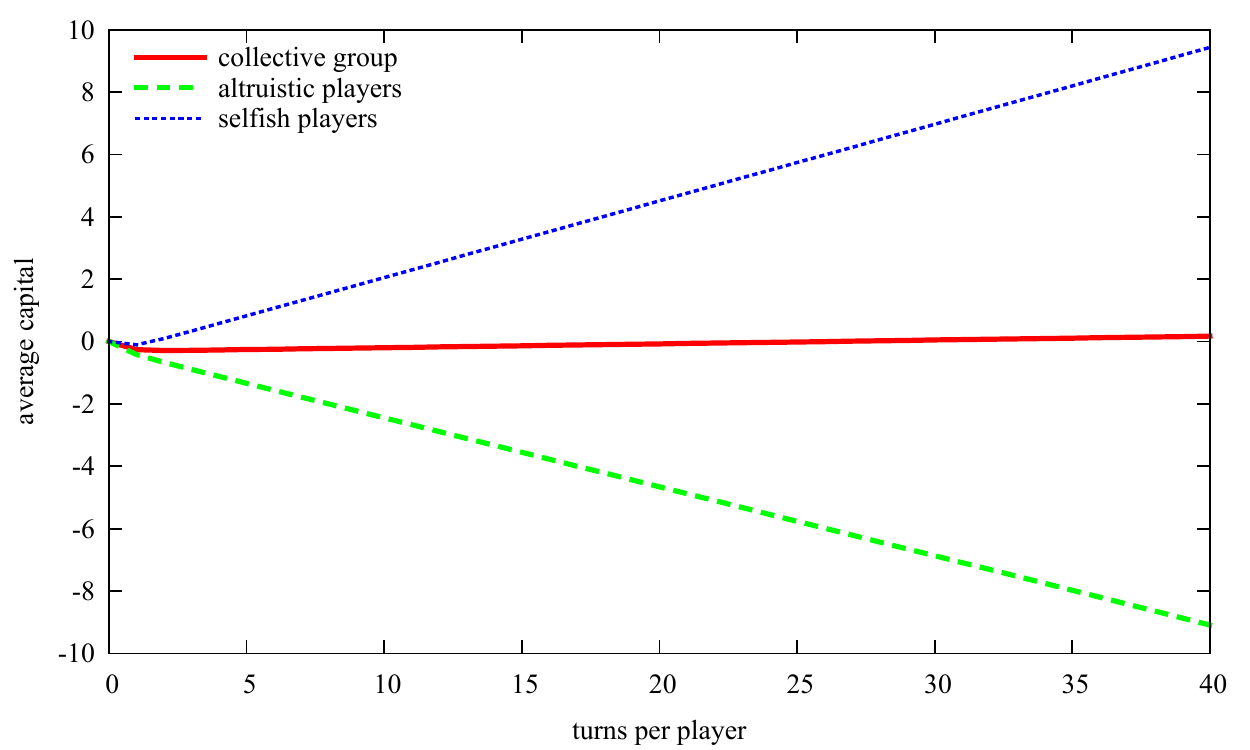}
\caption{{\bf AS model}. \Didasc Time evolution of the average capital of three groups: the collective group, altruistic players and selfish players. As usual, $\epsilon = 0.005$ and we have $N = 10,000$ players, half of which are altruistic and half are selfish. The results have been also averaged over 100 realizations, each one with $T=40$ turns per player.}
\label{differentiation-evol_naive}
\end{figure}

\begin{figure}[tbp]
\centering
\includegraphics[width=\columnwidth]{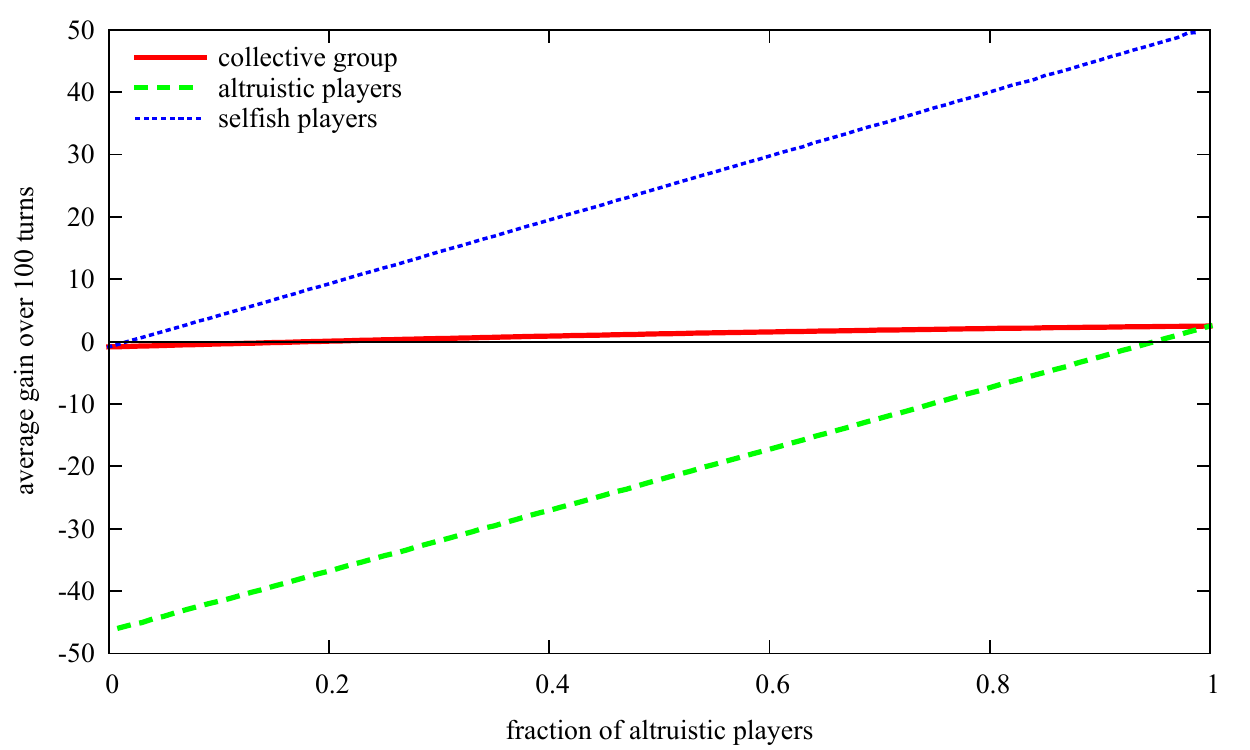}
\caption{{\bf AS model}. \Didasc Average gain over 100 turns as a function of the fraction $f$ of altruistic players. We report the average gain of the collective group, together with that of the two categories of players. The fraction $f$ varies from 0 to 1 with steps of 0.01. There are $N = 10,000$ players taking part in the game, with $\epsilon = 0.005$. The results are also averaged over 100 realizations.}
\label{differentiation-impact_naive}
\end{figure}

Once we have verified the positive effects of the presence of altruistic players, one could ask another question: while altruistic players give an advantage to the collective group considered as a whole (i.e. they produce an increase of the average capital per player), what happens to the two categories of players (selfish and altruistic ones) if we analyse them separately?
\\
To give a first answer to this question, we studied the case in which half of the players are altruistic and half are selfish (i.e. $N_{\mathrm a} = N_{\mathrm s} = N/2$, or in other terms $f=0.5$).
Then, we calculated the evolution of the average capital $\bar{C}(t)$ of the collective  group,  as well as the average capital of both altruistic and selfish players taken separately. The results are reported in Figure~\ref{differentiation-evol_naive}, where a total of $T=40$ turns per player have been considered.
The average capital of the collective group shows a small increase, of the order expected for $f=0.5$ on the basis of the results of Figure~\ref{fraction-capital_evolution}. But it clearly appears that this increment is a compromise between a sudden rise of selfish players' average capital and a specular decrease of altruistic players' one. This also means that the flow of capital from the altruistic group towards the selfish one clearly prevails on the average collective capital gain obtained from playing game~B.
\\
To better understand the equilibrium of capital between players, as a next step we calculated the average gain of capital $\bar{G}_C(f)$ over $100$ turns as function of different values of the fraction $f$ of altruists in the players community. For a given $f$, it is defined as $\bar{G}_C=\bar{C}(110) - \bar{C}(10)$, i.e. it simply consists in the difference between the average capital at turn 110 and at turn 10 (we started from time 10 to avoid the short initial transient of decreasing capital observed in Figures~\ref{fraction-capital_evolution} and \ref{differentiation-evol_naive}). We let $f$ vary from 0 to 1 and for each value of $f$ we calculated the corresponding average gain of the collective group and of the two categories of players.
We can take a look at the results in Figure~\ref{differentiation-impact_naive}.
%@@@@@@@@@@@@@
%@@@@@@@@@@@@@
%@@@@@@@@@@@@@
%@@@@@@@@@@@@@
\\
As we could expect from the results of Figure~\ref{fraction-capital_evolution}, the average gain of the collective group is slightly negative for small values of $f$ ($f<0.2$) and positive for high values of $f$. When $f = 0$, all the players are selfish ($N_{\mathrm s} = N$), therefore the average gain of the selfish group is exactly equal to the average gain of the collective group, i.e. slightly less than zero. In this case, we have no data for the gain of the altruists. On the other hand, when $f = 1$, all the players are altruistic ($N_{\mathrm a} = N$) and the positive gain of the altruistic group is equal to the gain of the collective group, while there is no data for selfish players.
For a wide range of intermediate values of $f$, we can see that selfish players have a big average gain (much bigger than the average gain of the collective group), whereas altruistic players have a big loss of the same order. The gain of the selfish increases with increasing values of $f$ because for large values of $f$ there are few selfish receivers surrounded by many altruistic contributors, so the capital received by each selfish player $S_i$ is big. On the other hand, when there are many altruists who randomly donate capital, a large fraction of the donations goes from an altruist $A_i$ to another altruist $A_j$, so the altruists have collectively a small loss of capital due to donations. We can see that, when $f$ is bigger than about 0.95, the altruists finally have a small positive gain.
\\
In order to better analyse this result, the gain of capital of the two categories of players can be thought as the sum of two distinct contributions. We know that the action of the altruists gives a collectively winning trend because, redistributing the capital, it lowers the value of $\pi_0$ and makes game~B profitable. But, apart from the general winning trend of game~B, there is a net movement of capital from altruistic players to selfish ones, due to game~A (played only by altruists). As a consequence, for a wide range of values of $f$, while the altruists have a small gain due to game~B (that remains profitable thanks to their actions), they have a much greater loss due to game~A (the altruistic action). On the other hand, selfish players have a small gain due to game~B (the same way as altruists have) and a big gain due to the donations of the altruists. To say it shortly, altruistic players favour the community at their own expense, while selfish players receive all the advantages without making any effort.

\section{Selective altruism: the SAS model}

\begin{figure}[tbp]
\centering
\includegraphics[width=\columnwidth]{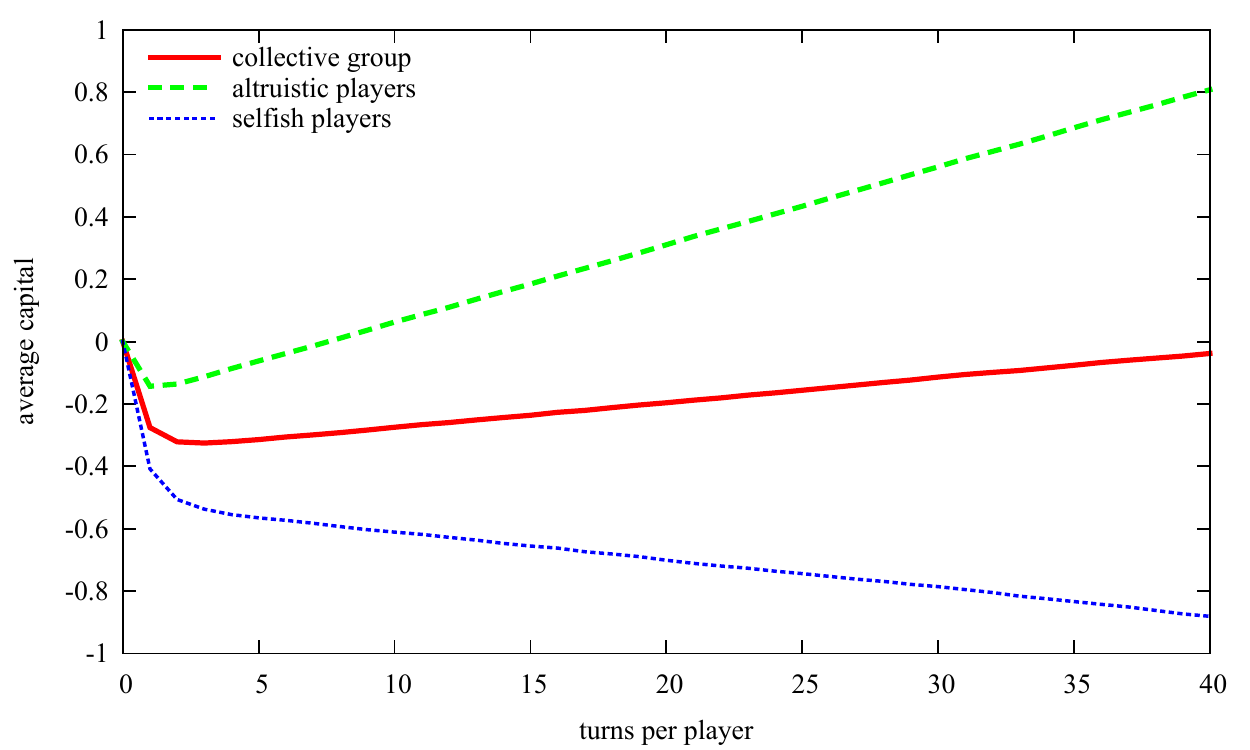}
\caption{{\bf SAS model}. \Didasc Time evolution of the average capital of three groups: the collective group, selective altruistic players and selfish players. We consider a community of $N = 10,000$ players, half of which are selectively altruistic and half are selfish. The results are averaged over 100 realizations. As usual, $\epsilon = 0.005$. }
\label{differentiation-evol_select}
\end{figure}

\begin{figure}[tbp]
\centering
\includegraphics[width=\columnwidth]{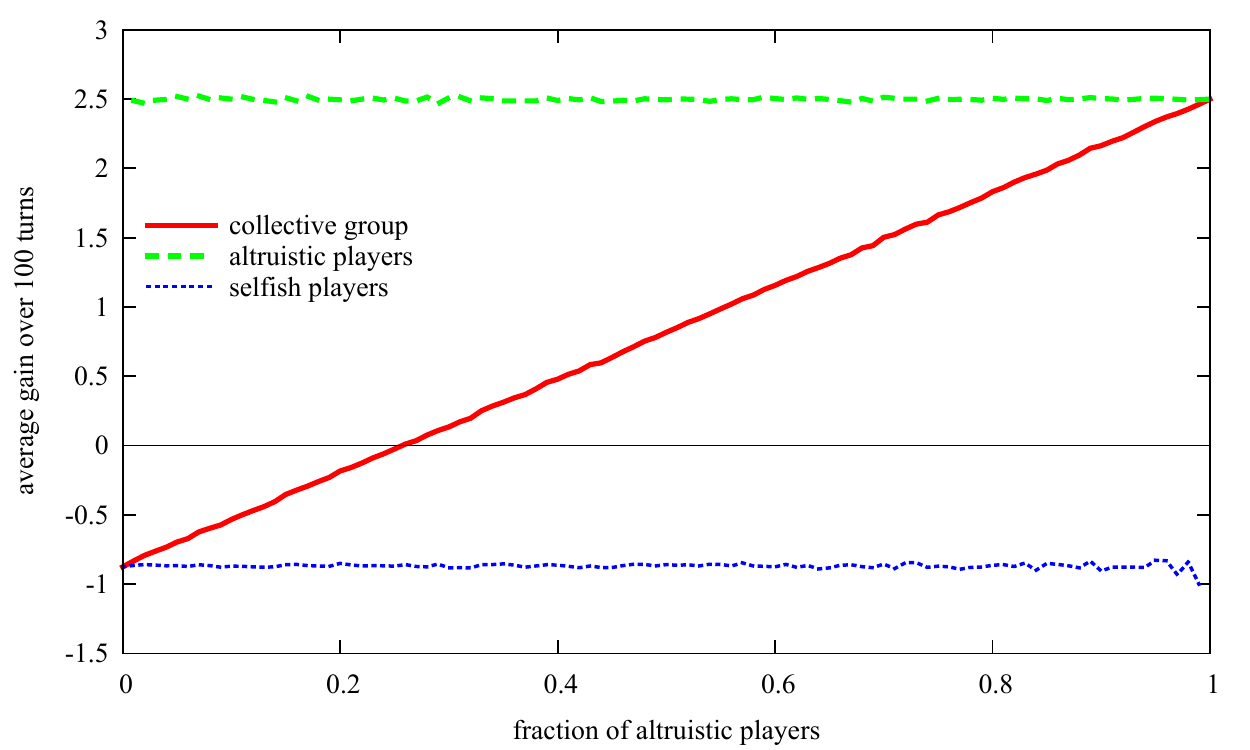}
\caption{{\bf SAS model}. \Didasc Average gain over 100 turns of the collective group, together with that of the two categories of players. The fraction of selective altruistic players is varied from 0 to 1 with steps of 0.01. There are $N = 10,000$ players that take part in the game, with $\epsilon = 0.005$. The results are averaged over 100 realizations.}
\label{differentiation-impact_select}
\end{figure}

The situation that we have just described seems therefore to encourage selfishness: the defection of just one player among many altruistic cooperators (i.e. $A_i \rightarrow S_i$) wouldn't almost affect the total redistribution of capital, but the newly selfish defector would have an individual drastic advantage. The problem is that if everyone follows this line of reasoning and turns into a selfish player, the community would end up with no altruists, which means that there would be no more capital movements, a situation that makes game~B not profitable, so ultimately everyone would lose capital. Is there a way to prevent this behaviour? Or, in other terms, is there another less naive way to be altruist, a way which favours the community without suffering personal disadvantages?
%@@@@@@@@@@@@@
\begin{figure}[tbp]
\centering
\includegraphics[width=0.8\columnwidth]{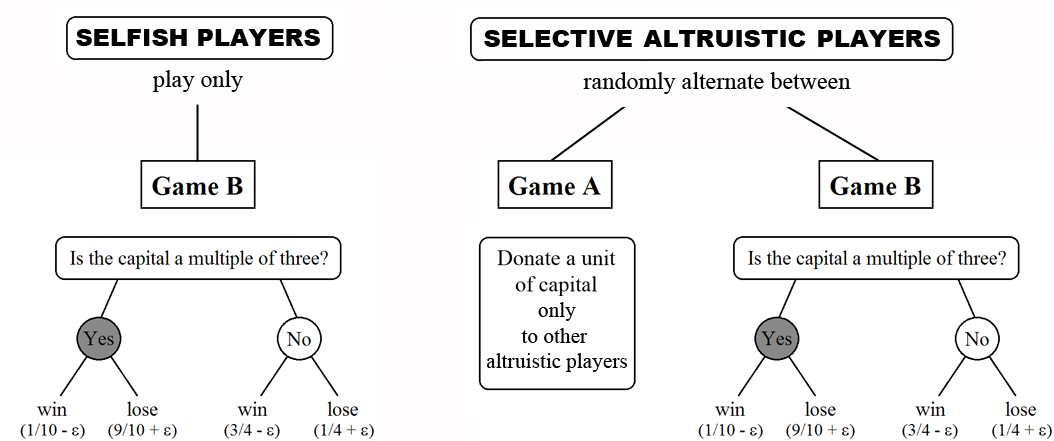}
\caption{{\bf SAS model}. \Didasc New rules for our collective model including the indirect reciprocity mechanism. Selfish players play only game B, while selective altruistic players randomly alternate between game A and game B, but they can give capital only to other selective altruistic players.}
\label{Selective_altruistic_rules}
\end{figure}
%@@@@@@@@@@@@@
\\
Among many types of collaborative behaviours, we found a particularly interesting mechanism called \emph{indirect reciprocity} that fosters the emergence of cooperation \cite{nowak1}. This mechanism occurs when an individual chooses whether to help or not another individual basing his decision on the reputation of the individual needing help. Those who have a reputation for giving aid to others who need it are more likely to receive help from a generous stranger when they find themselves in hard times.
In order to include the mechanism of indirect reciprocity in our model of fully interacting players, we restrict to the case in which everyone knows who is altruistic and who is selfish. In this case, who chooses to cooperate wants to do it only with other players who are cooperating. In practical terms, playing game~A, each altruistic player $A_i$ can give capital only to another altruistic player $A_j$.
That is what we call \emph{selective altruism}, in contrast to \emph{naive altruism} that we have seen in the previous sections. We will refer to this new model as {\it Selective Altruism-Selfishness (SAS) model} and we present it in Figure~\ref{Selective_altruistic_rules}. 
So, what are the results of such a behaviour? 
\\
To find out a first answer, we study the case in which half of the players are selectively altruistic and half are selfish ($N_{\mathrm a} = N_{\mathrm s} = N/2$). In this case, we calculate the evolution of the average capital of the collective group and that of the two categories of players. The results are displayed in Figure~\ref{differentiation-evol_select}.
Here we can see that the situation is drastically different from what we observed for AS model in Figure~\ref{differentiation-evol_naive}.
The average capital of the collective group continues to have a small positive trend, but now altruistic players have an increasing average capital, while selfish players have a decreasing capital.
Anyway, in both cases the overall variation is quite small, nearly ten times smaller than that one we observed in Figure~\ref{differentiation-evol_naive} for the capitals of the two categories.
It means that there is no flow of capital between the two categories and the variations of capital are now due only to game~B, therefore they are much slower than before.
%@@@@@@@@@@@@@
%@@@@@@@@@@@@@
\\
To better understand the case of selective altruism, we varied the fraction $f$ of altruistic players from 0 to 1 and calculated the corresponding average gains (over 100 turns) of the collective group and of the two categories of players.
The results of these calculations are presented in Figure~\ref{differentiation-impact_select}.
As we could expect, the gain of altruistic players is positive, whereas the one of selfish players is negative. The gains of the two categories remain almost constant in all the range of values of $f$ and, again, their absolute values are much smaller than the analogous ones in AS model shown in Figure~\ref{differentiation-impact_naive}.
\\
The explanation for this trend is quite simple. Selective altruists exchange capital only between them, creating a sort of network of mutual aid which does not communicate with the rest of the players. Independently of the number of altruists, their mutual exchange of capital ensures that game~B is profitable and gives to all of them the same probability of winning. So, the average gain of the altruists does not change with their number.
On the other hand, selfish players do not take part in the exchange of capital: they do not donate to anyone, but also do not receive from anyone. Now, each of them has a capital that changes only as a consequence of game~B, that we know to be losing if played singularly. So, all the selfish players have the same probability of losing and their negative average gain does not depend on their number.
In any case, there is no movement of capital from one category to the other and the only source of gain (or loss) for the average capital of the two categories is game~B.
That is why we do not see the great values of gain that we saw in the case of AS model.
%@@@@@@@@@@@@@
%@@@@@@@@@@@@@
Finally, we can see that the average gain of the collective group grows almost linearly since, for increasing values of $f$, there are more and more altruistic players in the community, who more and more positively influence the final outcome.

\section{Effects of imitation in AS and SAS models}

\begin{figure}[tbp]
\centering
\includegraphics[width=0.75\columnwidth]{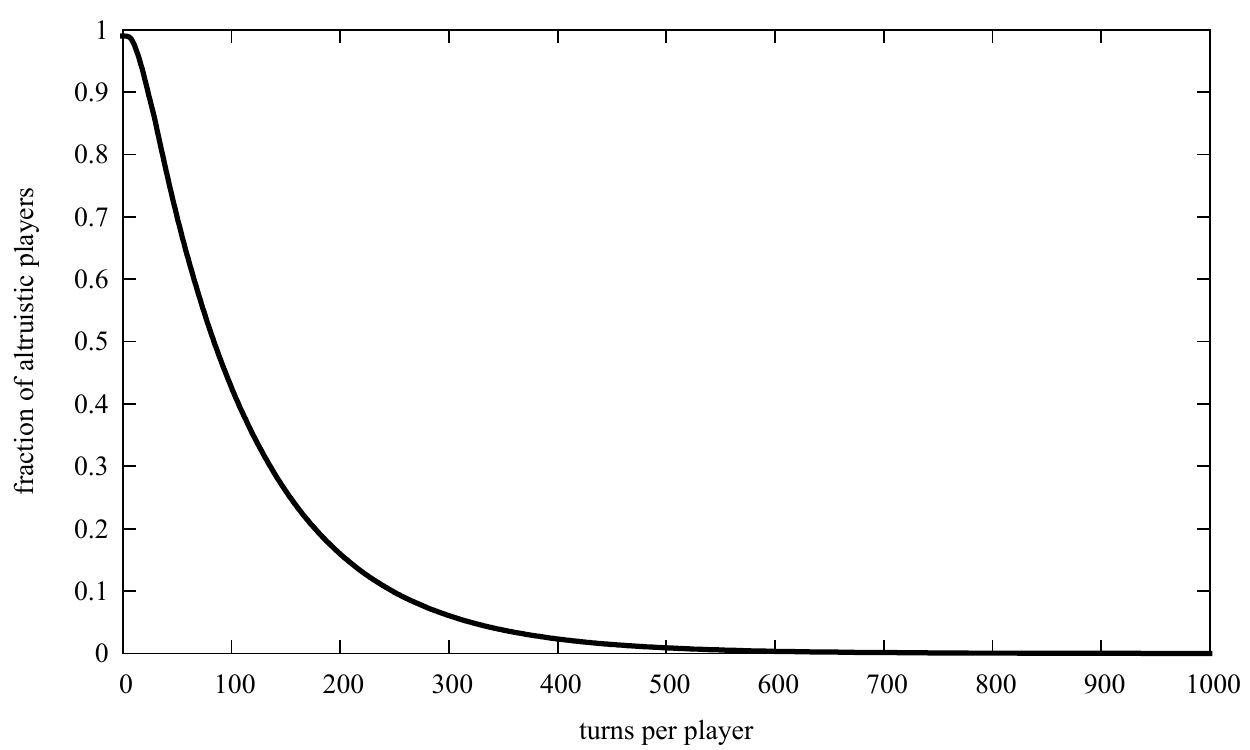}\\
\includegraphics[width=0.75\columnwidth]{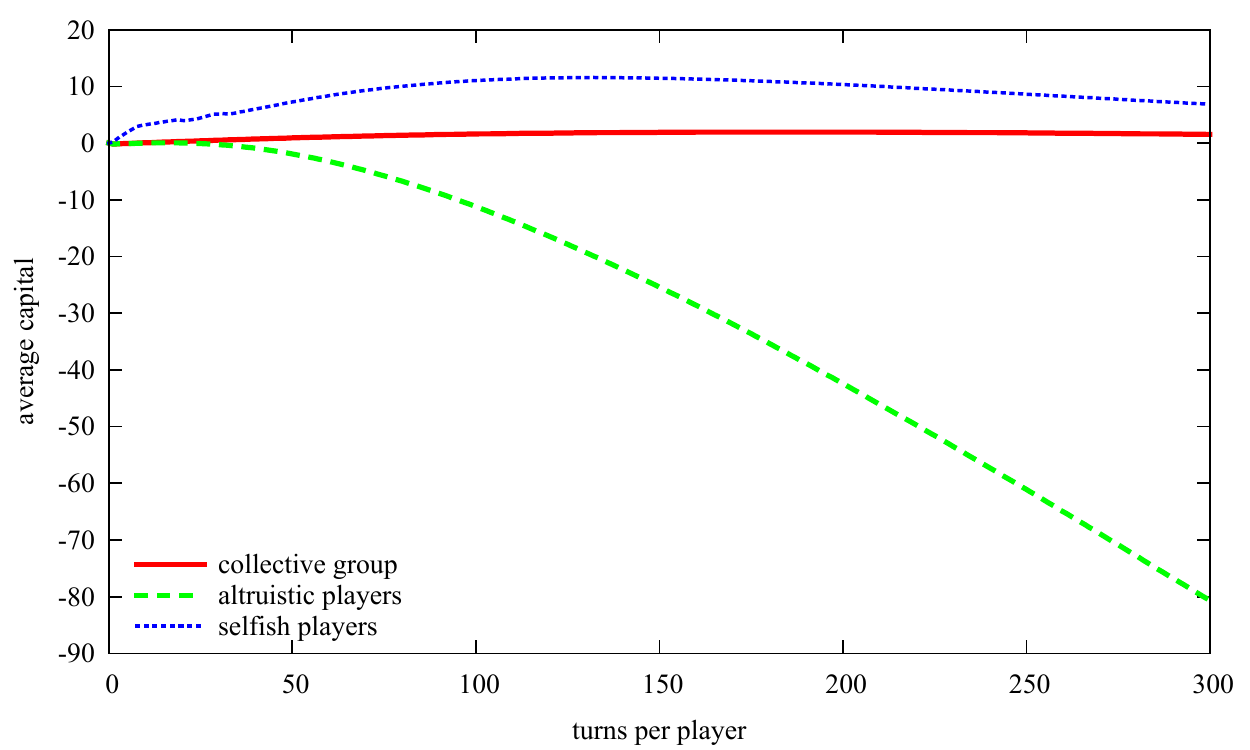}\\
\includegraphics[width=0.75\columnwidth]{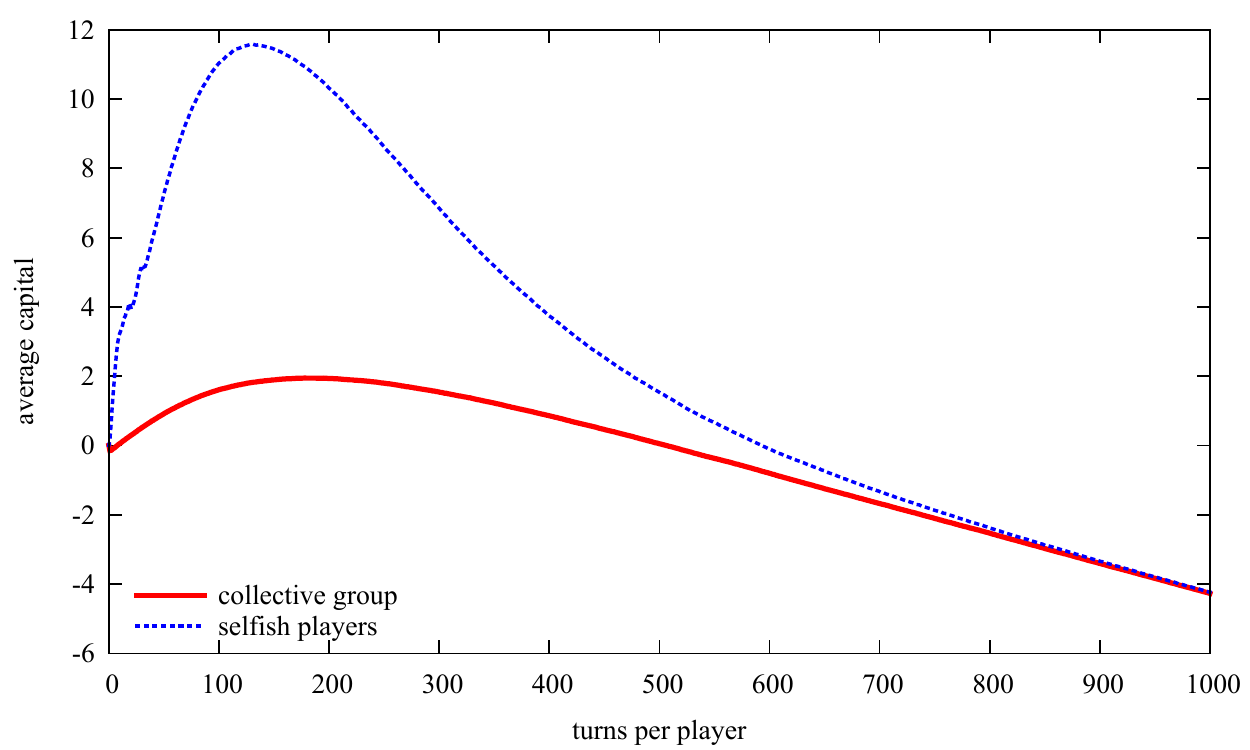}
\caption{{\bf Effects of imitation in AS model}. \Didasc Time evolution of a community composed of selfish players and of naively altruistic ones, where the new mechanism of imitation is present. We can see how the different capitals change over time, as well as the composition of the population. There are $N = 10,000$ players that take part in the game, with $\epsilon = 0.005$ and $\gamma = 0.01$. The results are averaged over 50 realizations. See text for further details.}
\label{diff_naive}
\end{figure}

\begin{figure}[tbp]
\centering
\includegraphics[width=0.75\columnwidth]{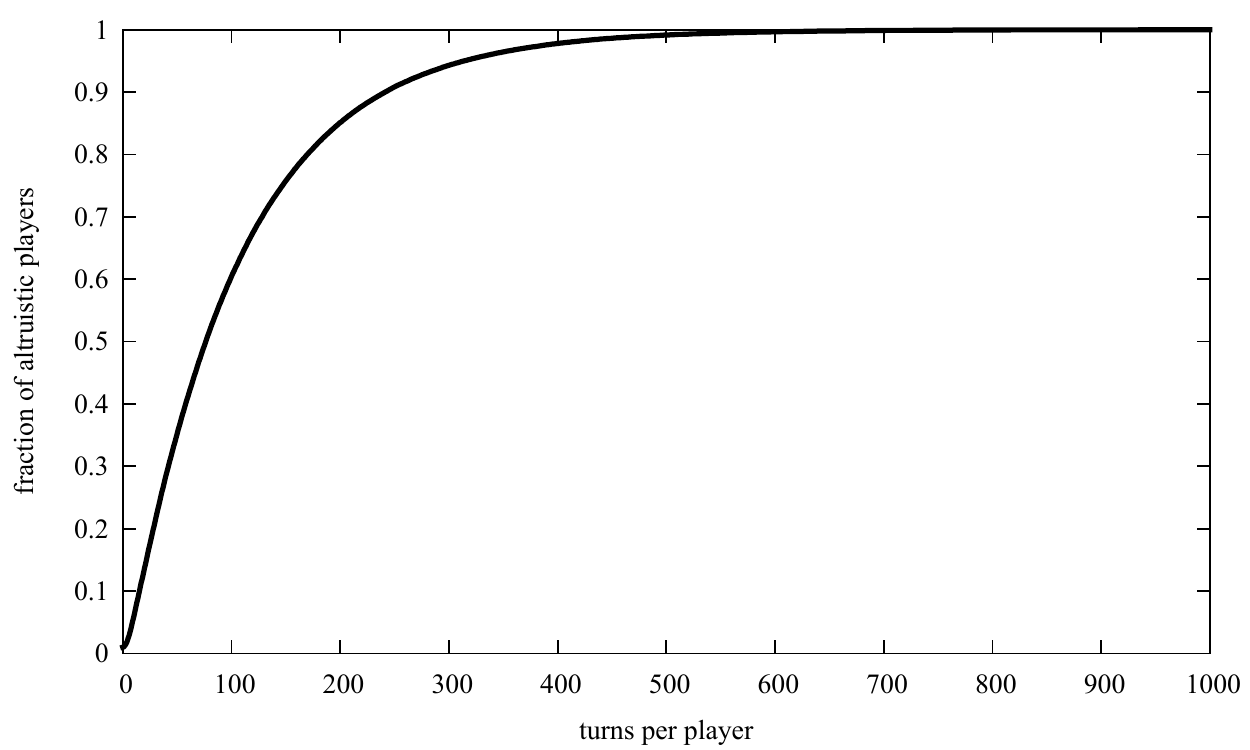}\\
\includegraphics[width=0.75\columnwidth]{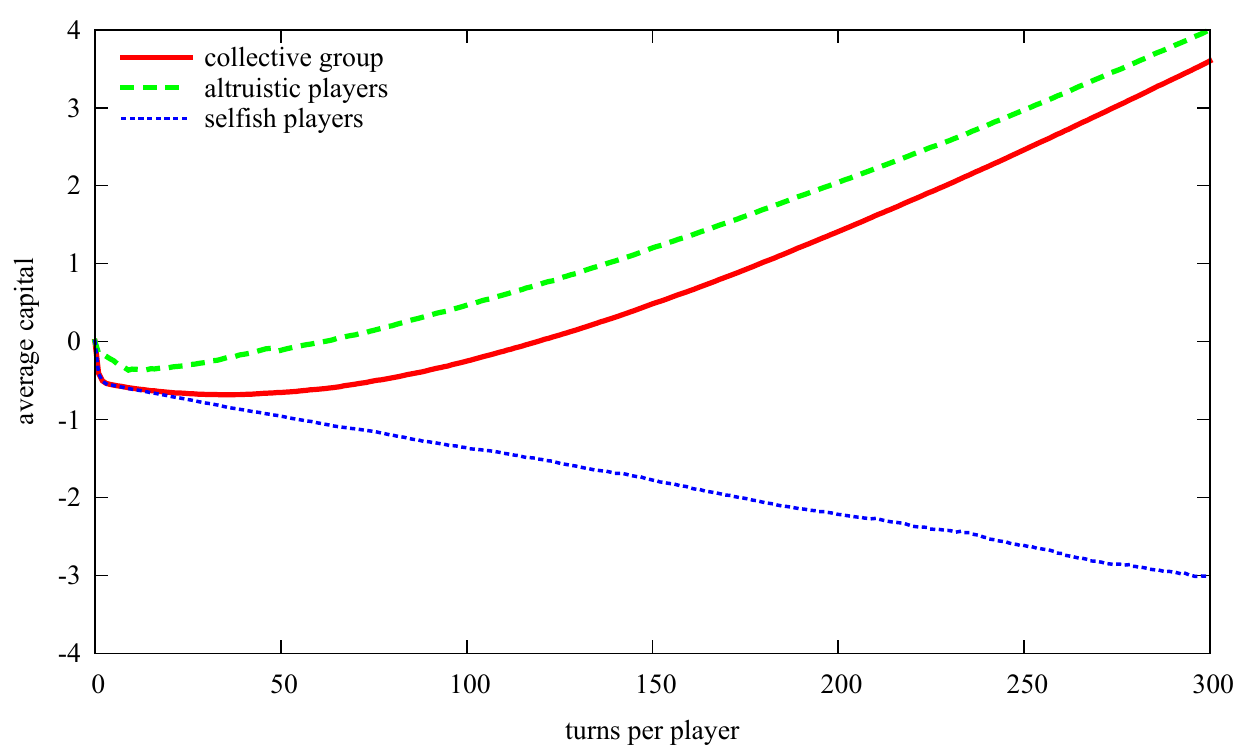}\\
\includegraphics[width=0.75\columnwidth]{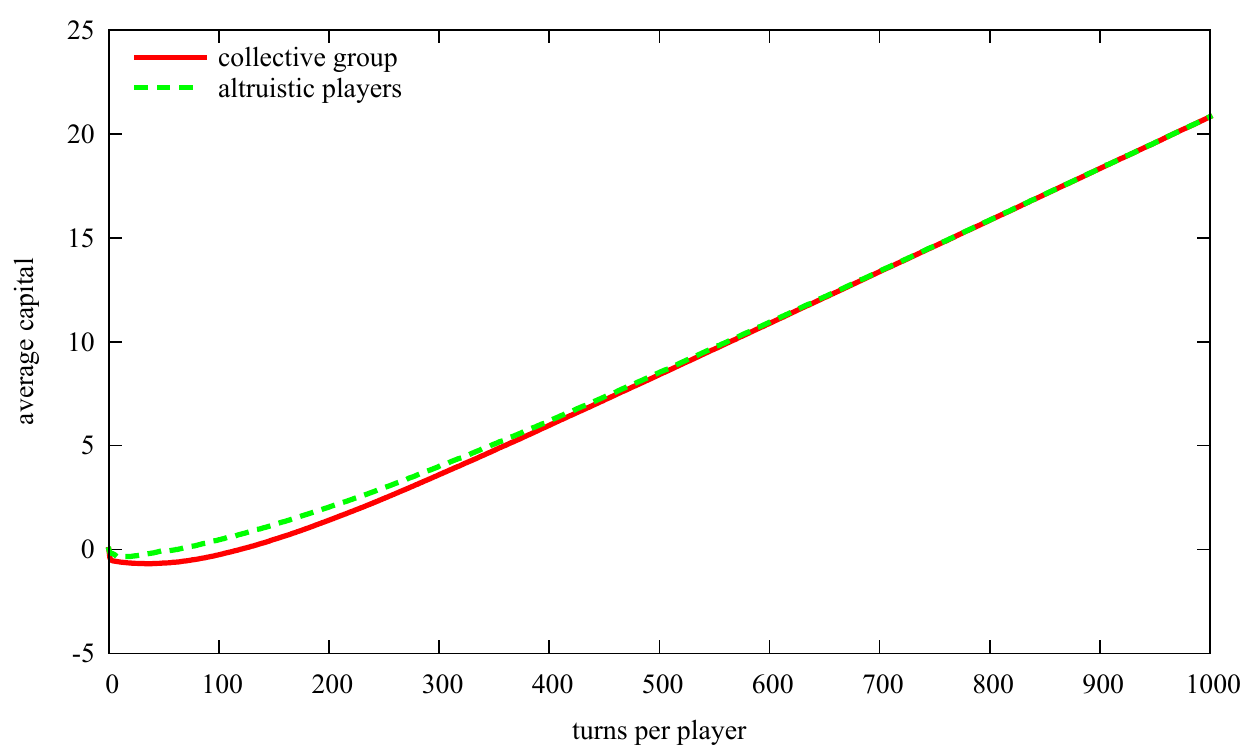}
\caption{{\bf Effects of imitation in SAS model}.  \Didasc Time evolution of a community composed of selfish players and of selectively altruistic ones, where the new mechanism of imitation is present. We can see how the different capitals change over time, as well as the composition of the population. There are $N = 10,000$ players that take part in the game, with $\epsilon = 0.005$ and $\gamma = 0.01$. The results are averaged over 50 realizations. See text for further details.}
\label{diff_select}
\end{figure}

Another interesting is  whether and how altruistic behaviour spreads among a population.
Why should people choose to help each other and cooperate rather than take advantage of possible altruists, pursuing their own profit?
In more detail, in a model with selfish and altruistic players and some mechanism by which players can change their own behaviour, we want to understand what could be the final result:
will one of the two categories prevail on the other one?
Or will there be a form of equilibrium between them?
\\
In \cite{helbing}, Helbing and Yu studied how social cooperation can arise spontaneously, based on local interactions rather than centralized control. They found that cooperation can survive or even emerge under adverse conditions, thanks to \emph{imitation} and success-driven migration. In that framework, the fundamental process is the organization of cooperators in clusters, which grant them higher payoff and allow them to survive and even grow in number thanks to imitation by other players.
Similarly, in \cite{nowak1} Nowak describes some mechanisms by which cooperation can emerge. Particularly interesting for our purposes is the so called \emph{spatial selection}: when cooperators form clusters they are more likely to grow and thus prevail in competition with defectors, whereas when cooperators and defectors are uniformly distributed the defectors usually win out. This is in total agreement not only with the results of Helbing and Yu, but also with the logic of \emph{group selection}: as Darwin himself wrote \cite{darwin}, ``a tribe including many members who {\ldots} were always ready to aid one another, and to sacrifice themselves for the common good, would be victorious over most other tribes; and this would be natural selection''.
\\
Inspired by these findings, we further modified our model including also the possibility for a player to change his own behaviour by means of imitation, i.e. through some kind of herding mechanism. At each time step, a given player $P_i$ will be randomly selected as in the previous simulations. Now, however, before playing his turn according to his own nature (altruistic or selfish), player $P_i$ will look at all the others and will find the richest one, say $P_j$: if player $P_j$ is richer than player $P_i$, player $P_i$ will adopt with probability $\gamma = 0.01$ the strategy (i.e. the altruistic or selfish behaviour) of player $P_j$. As a consequence of this mechanism, the composition of the population will change over time and our aim is precisely to study this evolution.
\\
First, let us take into consideration the case of AS model, where altruistic players are naive.
We already know that naive altruism is a loosing strategy in terms of capital, so we expect that selfish behaviour will spread throughout the whole community.
Therefore, we set the initial conditions so that 99\% of the population is altruistic and only 1\% is selfish.
The results of the simulations are shown in the three panels of Figure~\ref{diff_naive}.
%@@@@@@@@@@@@@
%@@@@@@@@@@@@@
\\
The plot in the top panel immediately confirms our expectations. In fact, the fraction of altruists steadily decreases in time from its initial value, 0.99, and asymptotically goes to zero, while selfishness rapidly ``infects'' the entire population. This also means that in this model, where people tend to imitate the richest players, naive altruism has no possibility to survive.
\\
In the middle panel we see the corresponding initial evolution (until turn 300) of the average capital, calculated for the whole community and, separately, for the two categories of players. The altruists' capital is initially almost constant (actually, looking at numerical data, one can  appreciate a slight increase), but very soon starts to decrease and then maintains a negative slope. That is because in the first turns there are few selfish players draining the capital of altruistic players, but the selfish group starts soon to grow in number and so altruistic players lose capital with their donations. On the other hand, selfish players have from the beginning an increasing capital, because they can enjoy all the advantages of the altruists presence. They are successful because they gain capital at the expense of altruists, so they are imitated by poorer altruists and grow in number. However, at around turn 150, altruistic players become too few to ensure selfish players an increasing capital, therefore selfish players start to lose capital too. Consequently, also the average capital of the collective group follows a similar evolution and, after an initial small increase, it starts to decrease.
\\
If one wants to see the global evolution of the capital over the entire simulation, this is plotted in the bottom panel. We stopped at turn 1000 because after that there are no substantial changes. In this panel we do not plot the evolution of capital of altruistic players, because after turn 300 their number becomes too low to have a stable average over the different realizations. We can see that the decreasing trend of both the collective group capital and the selfish group capital continues well after turn 300.
Finally, at about 800 turns, the two averages tend to coincide, as we  expect for a population made up entirely of selfish players.
\\
Summarizing, we have found that, at the beginning, naive altruism seems favourable for the entire community, even if unfavourable for the altruists, because who defects and acts selfishly gets the most advantages. So selfish players are seen as a winning model and get imitated, causing a spread of selfish behaviour. In such a way, everyone will eventually become selfish and care only about himself. However, none of the players can win if playing alone. Therefore, all the players end up losing capital --- the worst situation for the whole community. Selfishness, even if initially profitable at the individual level, ultimately turns out not to be a winning strategy at the global level.
\\
After having studied the outcome of competition between selfishness and naive altruism in AS model, as a second step we took again into consideration SAS model, where selective altruists donate capital only to other altruists.
We wanted to check whether such a more robust form of altruism  spreads or not throughout the entire community, so we performed a new set of simulations starting this time from an initial condition with 99\% of selfish players and 1\% of selectively altruistic ones.
The results are plotted in the three panels of Figure~\ref{diff_select}.
%@@@@@@@@@@@@@
%@@@@@@@@@@@@@
\\
In the top panel we immediately see that the situation is the opposite of the previous one: now it is the altruism that rapidly diffuses among the population. In fact, starting from the value of 0.01, the fraction of altruistic players strongly increases and then asymptotically tends to 1. Selective altruism seems, therefore, the preferred behaviour and for this reason it is imitated by the majority of the players.
\\
In the middle panel we zoom on the first 300 turns of time evolution of the three average capitals of interest. Selfish players' capital presents an evident decrease, that stays almost constant along all the considered interval. It happens because they are now cut away from the network of capital exchange: they do not gain by altruists donations and, playing only game~B, they are stuck in a losing trend. On the other hand, altruistic players have a small gain from the beginning, because even when they are in a small number they manage to exchange capital between them (but only between them) and obtain a positive effect on game~B. Mutual aid gives an advantage to altruistic players and makes them almost immediately richer than selfish ones, so the former are regarded by other players as a winning model and are imitated, growing in number. The average capital of the collective group has another trend: it initially decreases and then, around turn 40, starts to increase. That is because initially almost all the players are selfish and lose capital, but when more and more players become altruistic and have an increasing capital, collective capital also starts to grow.
\\
Finally, in the bottom panel, we plot the whole evolution of the system. Again, we stopped at turn 1000 because there are no significant changes after that. Selfish players' capital is not plotted for the same reason as before, because after a certain time there is not a sufficient number of them. It can be seen that collective and altruists capitals continue their increasing trend. From the beginning, selfish players join the altruistic faction, but being poorer than the players they join, they initially slow the increase of average capital of the altruistic group. At last, when almost all the players are altruistic, the two average capitals become equal and the increasing trend has a constant slope.
\\
To summarize the situation with different words, we have seen that selective altruism is the winning behaviour: this kind of altruists can help each other and thus gain capital, without being ``robbed'' by the selfish. Besides, they give a positive image to selfish players, who sooner or later decide to follow their example and have a kind of ``social improvement'', becoming altruistic. In the end, all the players help each other and have a positive trend of increasing capital --- the ideal situation for the whole community.

\section{Exploring different topologies}

\begin{figure}[tbp]
\centering
\includegraphics[width=\columnwidth]{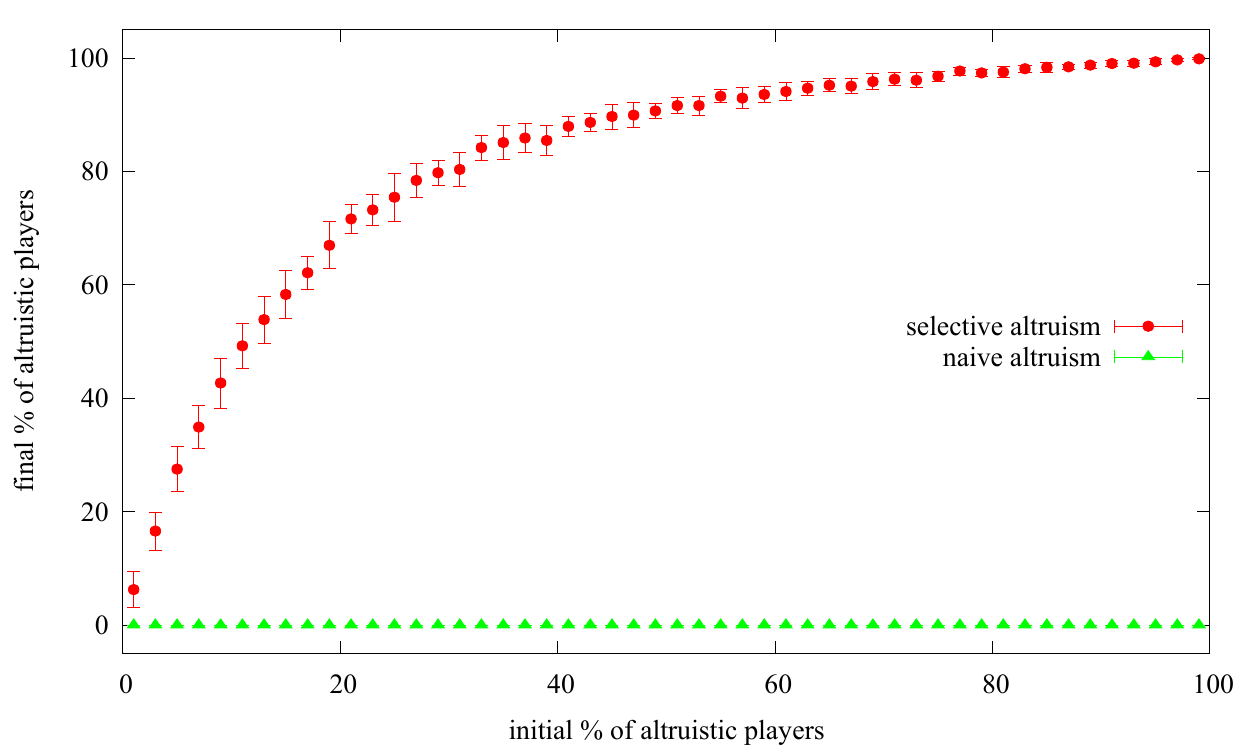}
\caption{{\bf Regular lattice}. \Didasc Final percentage of altruists (taken at time 10,000) as a function of their initial percentage.
We consider AS and SAS models with imitation.
There are $N = 1936$ players, which can be selfish or (selectively or naively) altruistic, arranged in a square lattice.
The results are averaged over 20 realizations, so we report the averaged values together with bars indicating their standard deviations.}
\label{plot_latt_sel}
\end{figure}

In the previous sections we always considered a  fully interacting community of altruistic and selfish players $\{P_i\}_{i=1,2,\ldots,N}$, where each of them  interacts with every one else. In this section we want to explore what happens if a given player $P_i$ can interact only with a limited number of neighbours, like in a  real social network (in this respect, the fully interacting community is equivalent to an all-to-all network). In other words, for both AS and SAS models with imitation, we study here the influence of the network topology on the evolution of cooperation in our community of players and on the emergence of different patterns of collective behaviours. In practice, once selected a given network topology, a player $P_i$ who plays game~A can, now, donate capital only to a player $P_j$ belonging to his neighbourhood $\{P_j\}_{j=1,2,\ldots,k_i}$ (being $k_i$ the degree of player $P_i$), and the same rule holds for imitation: each player can look only at his neighbours and eventually imitate the strategy of the richest one among them.
\\
%@@@@@@@@@@@@@@@@@@@@@@@@@
\begin{figure}[tbp]
\centering
\subfloat[][Regular]
{\includegraphics[width=.45\columnwidth]{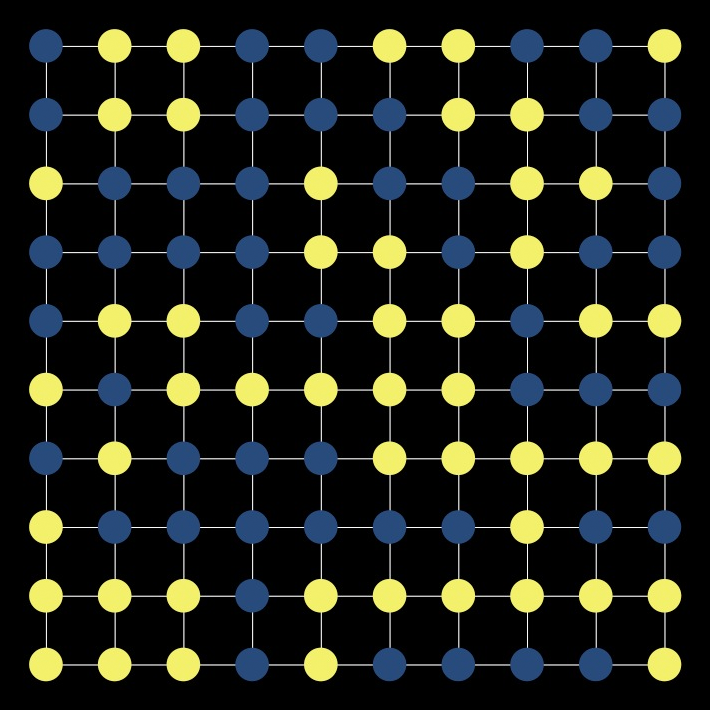}} \quad
\subfloat[][Small-World]
{\includegraphics[width=.45\columnwidth]{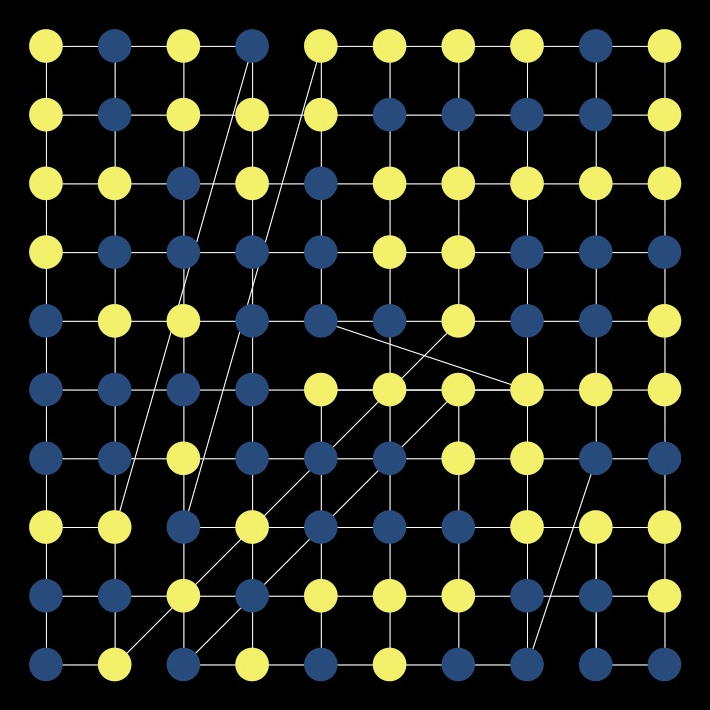}} \\
\subfloat[][Random]
{\includegraphics[width=.45\columnwidth]{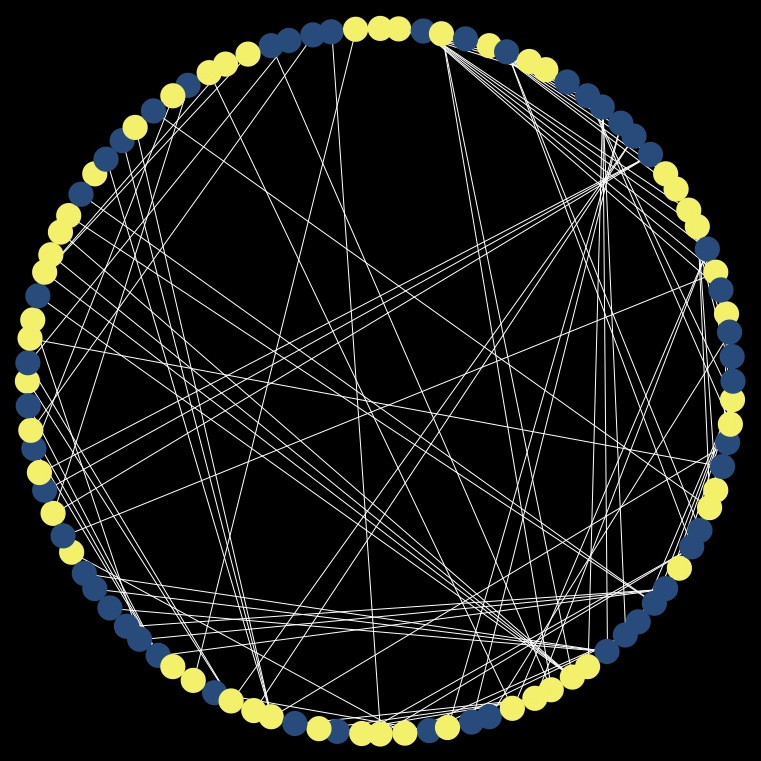}} \quad
\subfloat[][Scale-Free]
{\includegraphics[width=.45\columnwidth]{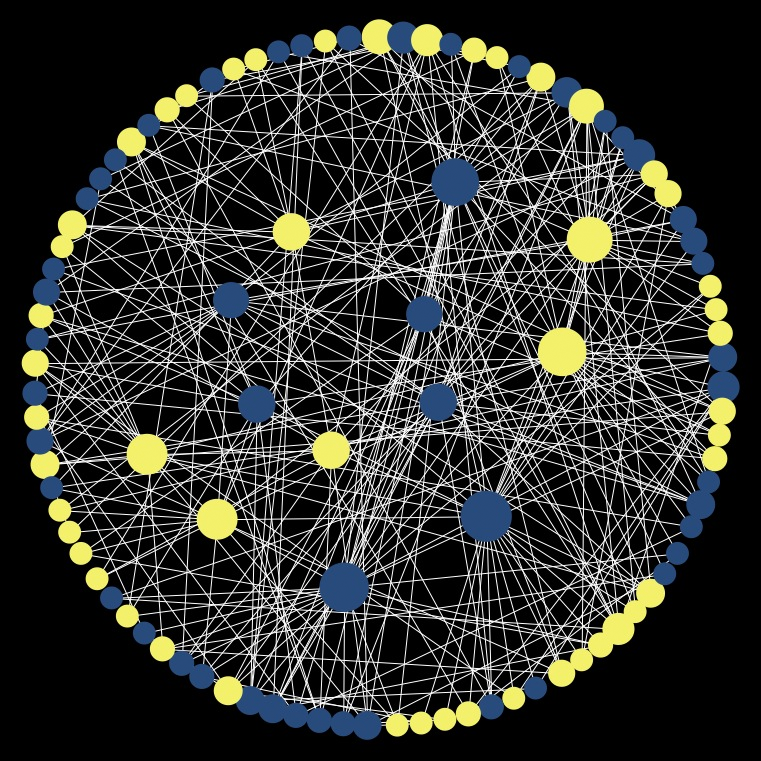}}
\caption{{\bf Different topologies}. \Didasc In the four panels we report four examples of the network topologies we adopt in this section. For a better visualization, we show networks with $N=100$ nodes, while in the simulations we considered  $N=1936$. Blue nodes represent selfish players, while yellow nodes represent altruistic ones.} 
\label{networks}
\end{figure}
%@@@@@@@@@@@@@@@@@@@@@@@@@
To understand the importance of topology in this collective game, we tested our model over four different types of networks, that can be seen in Figure~\ref{networks}.
\begin{itemize}
\item The \emph{regular lattice} is, in our case, an ensemble of players arranged like a 2-dimensional square lattice (with open boundary conditions), where every site is connected only to its four nearest neighbours.
\item The \emph{small-world network} is obtained from the regular lattice, rewiring at random every pair of links with a probability $p = 0.02$. It has two remarkable properties: a small distance between nodes (like a random network) and a high clustering coefficient (like a regular lattice) \cite{watts-strogatz}.
\item The \emph{random network} is created  by adding at each step a new node, connected to a randomly chosen old node (it is often represented as a circle).
\item Finally, the \emph{scale-free network} is created by a mechanism called ``preferential attachment'': at each step a new node is added and is connected to the existing nodes with a probability proportional to their degree. This mechanism produces a degree distribution that follows a power law \cite{barabasi} (we again represent the network as a circle and  the most connected nodes are drawn inside it).
\end{itemize}
%@@@@@@@@@@@@@@@@@@@@@@@@@
\begin{figure}[tbp]
\centering
\subfloat[][$t = 0$]
{\includegraphics[width=.31\columnwidth]{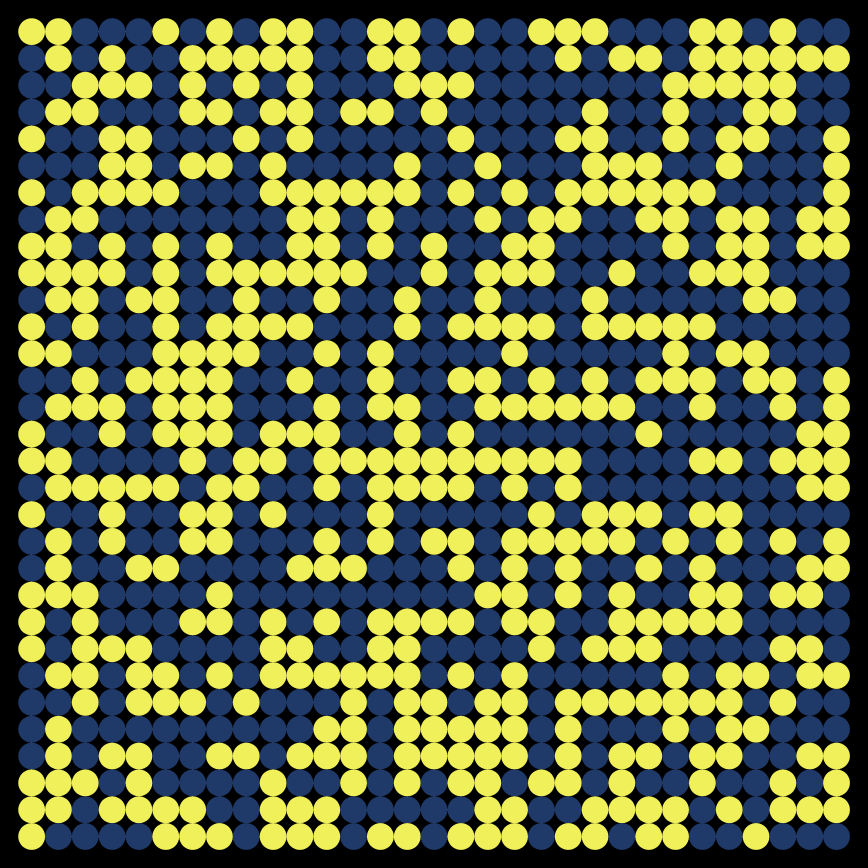}} \quad
\subfloat[][$t = 2$]
{\includegraphics[width=.31\columnwidth]{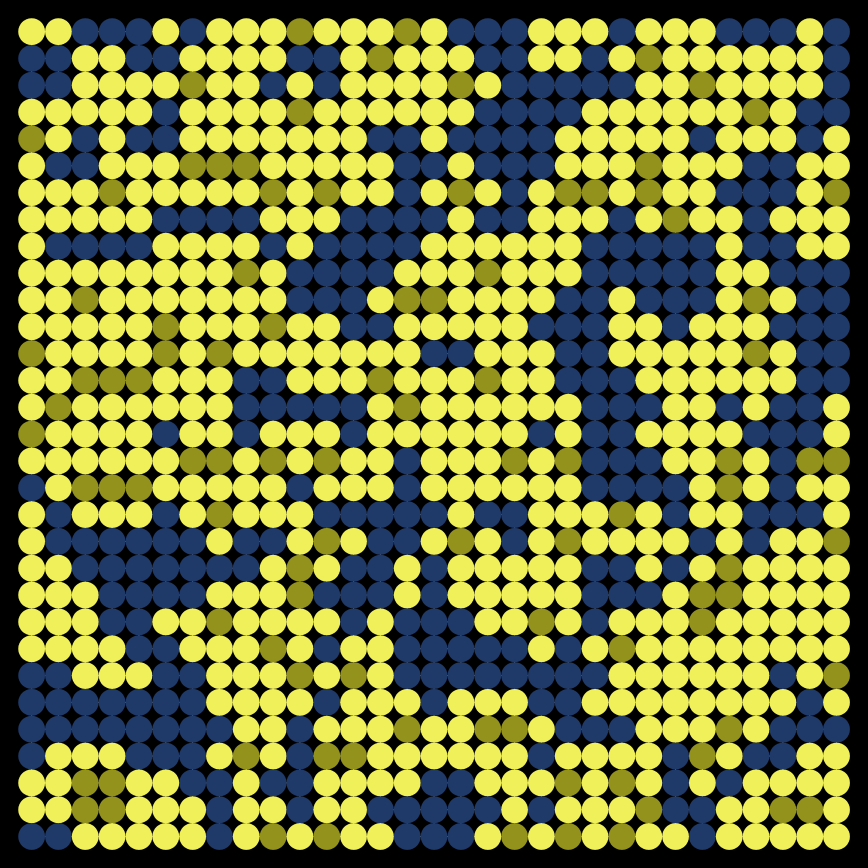}} \quad
\subfloat[][$t = 10$]
{\includegraphics[width=.31\columnwidth]{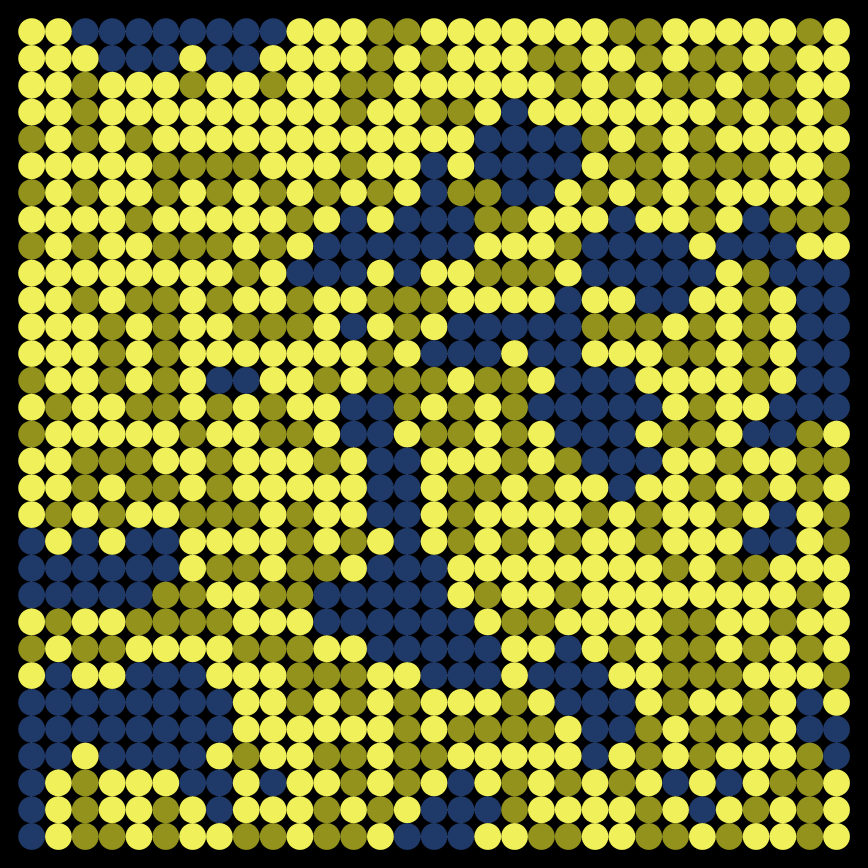}} \\
\subfloat[][$t = 100$]
{\includegraphics[width=.31\columnwidth]{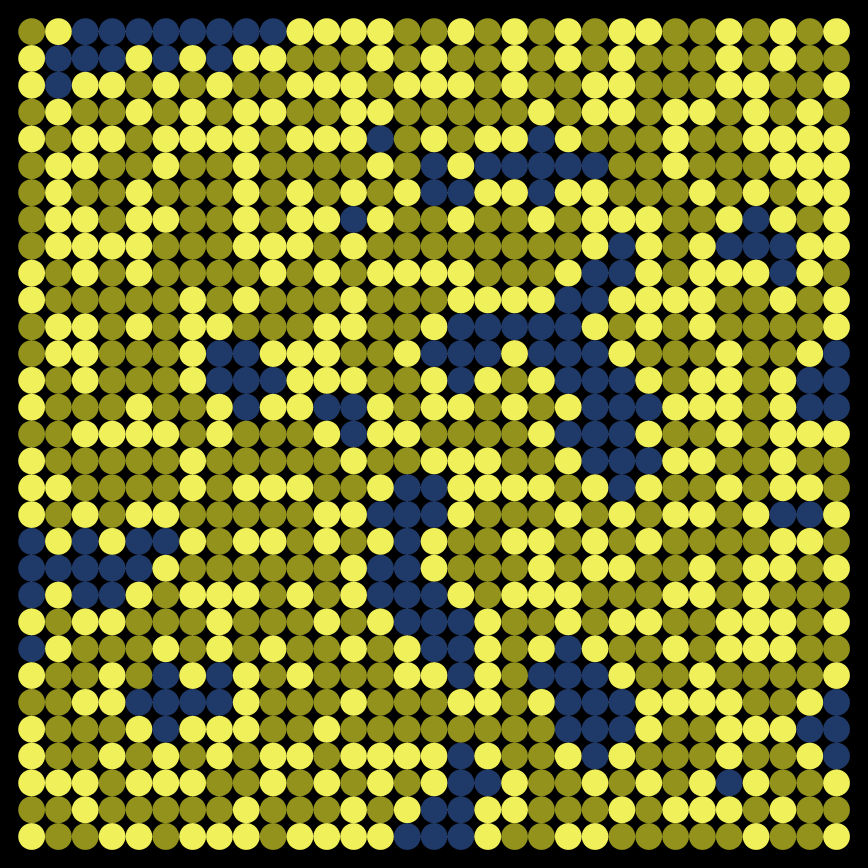}} \quad
\subfloat[][$t = 1000$]
{\includegraphics[width=.31\columnwidth]{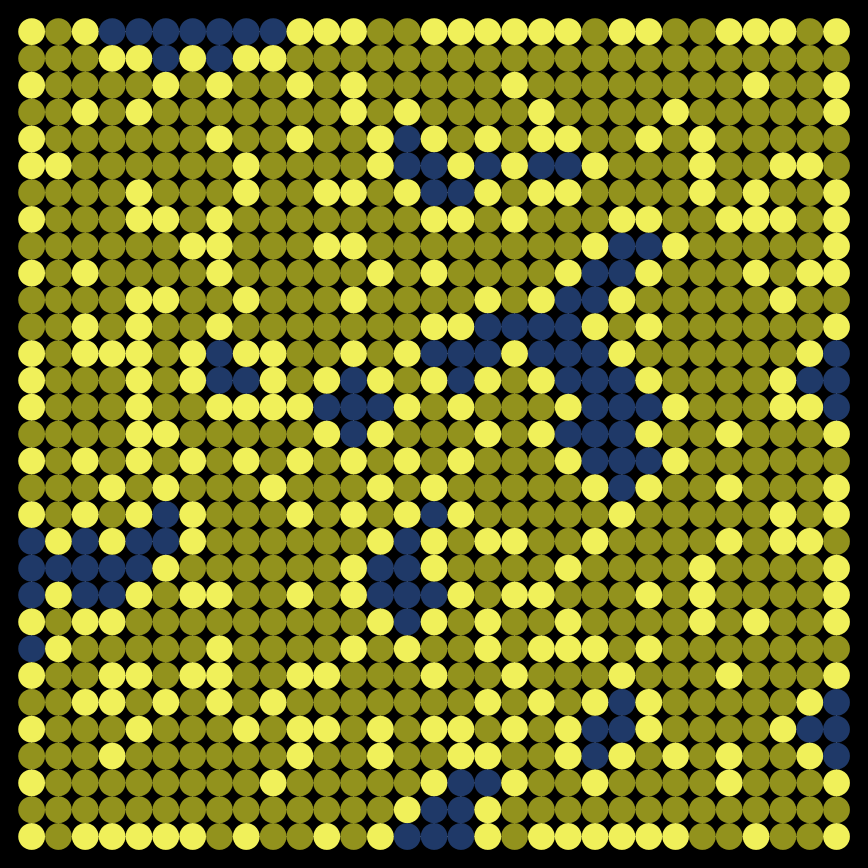}} \quad
\subfloat[][$t = 10000$]
{\includegraphics[width=.31\columnwidth]{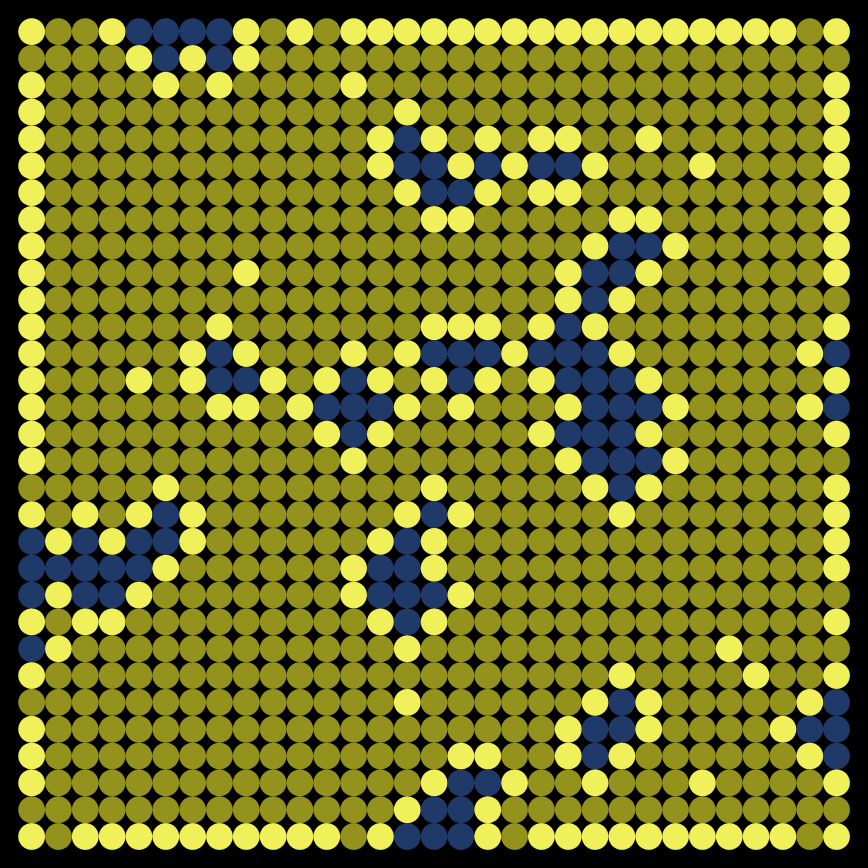}}
\caption{{\bf Regular lattice}. \Didasc Graphic representation of the evolving ensemble of players on a regular lattice topology for SAS model.
For a better visualization, we adopt a network with $N=961$ players portraied at 6 different times (turns per player).
At $t = 0$, one half of the population (chosen at random in the lattice) is selfish, whereas the other half is selectively altruistic.
Colours have the same meaning as in Figure~\ref{networks}.
For yellow (altruistic) nodes, a dark shade of colour means that the capital of the player is positive, whereas a light shade indicates a null or negative capital.
}
\label{latt_evol_snapshots}
\end{figure}
%@@@@@@@@@@@@@@@@@@@@@@@@@
In the following, the size of each considered network will always be of $N=1936$ players (even if, for a better visualization, the sizes represented in the figures will be usually smaller).
Besides, we made another change to our model: to have a faster evolution, we set $\gamma = 1$ in the herding process, meaning that imitation is certain if conditions on capital are fulfilled. 
\\
Let us start by considering SAS model in the regular lattice.
Given an initial population of selfish (blue nodes) and selective altruistic (yellow nodes) players, we observed that its composition evolves during time and asymptotically reaches a final stable condition with some interesting emergent patterns. Just to give an example of this process, in Figure~\ref{latt_evol_snapshots} we report some subsequent snapshots of a single simulation run: the sudden evolution at the very first instants of time becomes slower and slower and the system tends to reach a steady state, where the (selfish or altruistic) strategy of the players does not change any more. Actually, if we compare panels (e) and (f), we can see that the colour composition is almost the same but, on the other hand, a visible difference still holds about capital: in fact, many altruistic poor players (light yellow) have become richer (dark yellow), and the surviving poor altruists tend to surround clusters of selfish players or to occupy the borders of the lattice.  
\\
It is interesting to explore in deeper detail these stationary emerging patterns of cooperation. In order to do this, we varied the initial percentage of altruistic players from 1\% to 99\% and, for each value, we calculated the corresponding value of the final steady percentage of altruists for both AS and SAS models.
We repeated each simulation over 20 different events and report in Figure~\ref{plot_latt_sel} the averaged values of this quantity together with its standard deviation.
%@@@@@@@@@@@@@@@@@@@@@@@@@
%@@@@@@@@@@@@@@@@@@@@@@@@@
The results show that, in the case of naive altruism, the final percentage of altruists is always zero --- the same result that we found for the all-to-all network.
On the other hand, in the case of selective altruism, plotted data form a peculiar curve, meaning that the final percentage of altruists strictly depends on the initial percentage. This is a completely different situation with respect to the all-to-all network, where the final outcome (an entirely altruistic population) was independent of the initial composition of the population. Introducing a lattice topology has, evidently, the effect of preventing a complete diffusion of selective altruism, creating a level of saturation that depends on the initial composition. Furthermore, as we have seen in the previous figure, the final stationary state of the system presents interesting patterns of altruism-selfishness. 
\\
In order to better understand the origin of these patterns in SAS model, we report in Figure~\ref{snapshots} the final composition of the population for four different values of the initial percentage of selective altruists in a regular lattice with a smaller number of nodes (961), for a better visualization. 
Besides, in Figure~\ref{enlarg}, we show six enlargements of some significant patterns emerging from the previous figure.
%@@@@@@@@@@@@@@@@@@@@@@@@@
\begin{figure}[tbp]
\centering
\subfloat[][5\%]
{\includegraphics[width=.31\columnwidth]{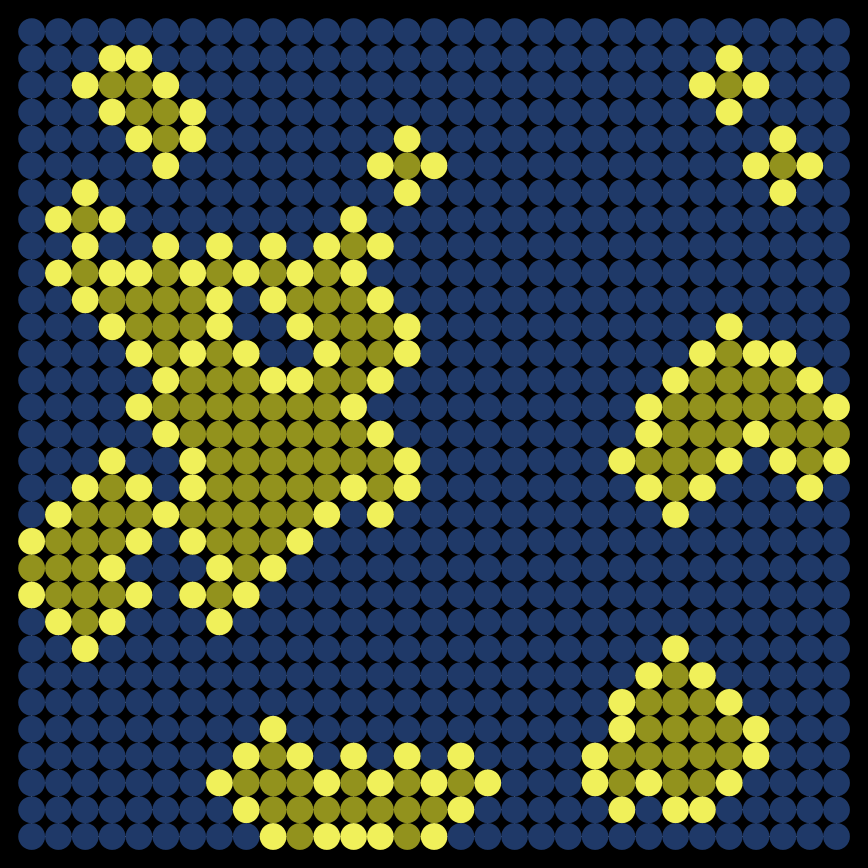}} \quad
\subfloat[][10\%]
{\includegraphics[width=.31\columnwidth]{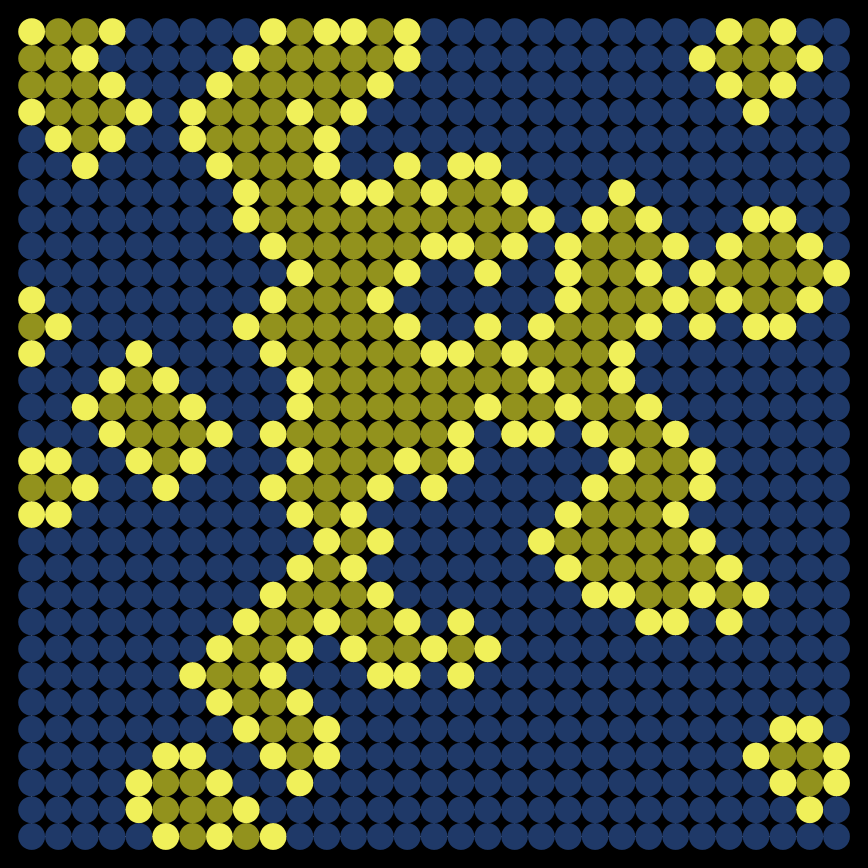}} \\
\subfloat[][20\%]
{\includegraphics[width=.31\columnwidth]{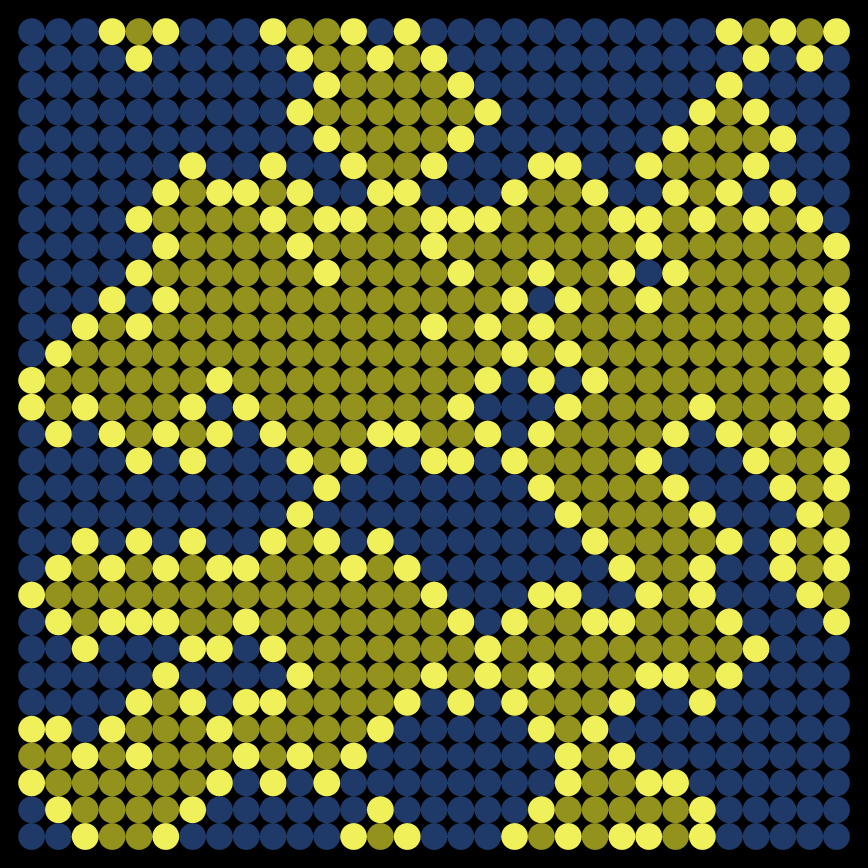}} \quad
\subfloat[][50\%]
{\includegraphics[width=.31\columnwidth]{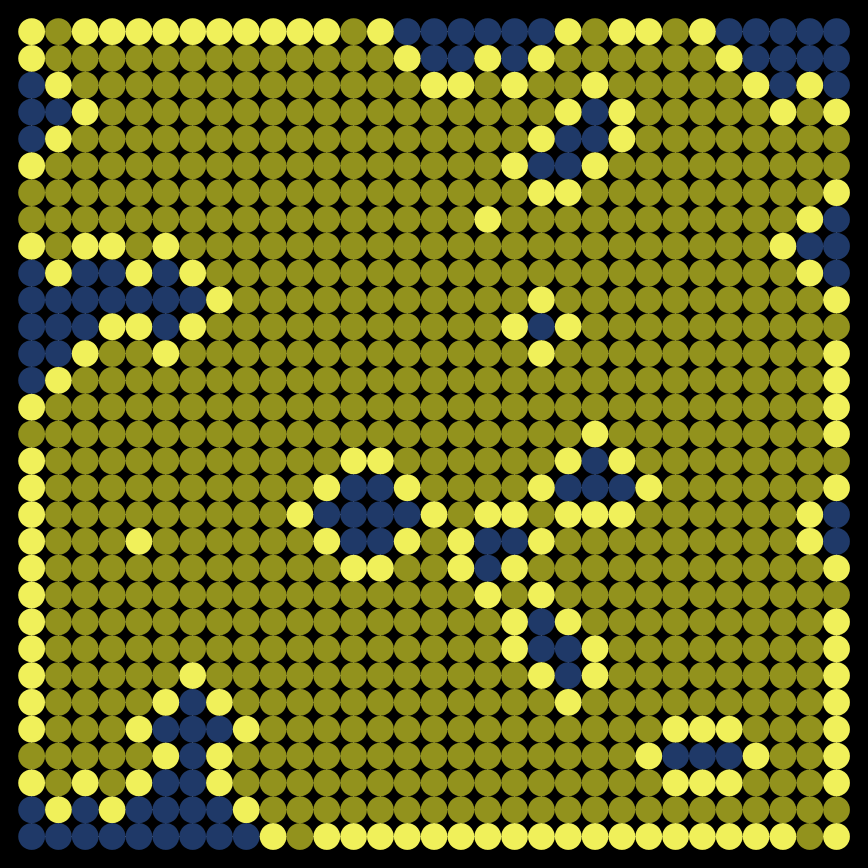}}
\caption{{\bf Regular lattice}. \Didasc In the four panels we report four examples of the final composition of population (at time 10,000) obtained from four initial percentages of altruists (5\%, 10\%, 20\% and 50\%) in SAS model.
There are 961 players, which can be selfish or selectively altruistic, arranged in a square lattice. Colours have the same meaning as in Figure~\ref{latt_evol_snapshots}.}
\label{snapshots}
\end{figure}
\begin{figure}[tbp]
\centering
\subfloat[][]
{\includegraphics[width=.20\columnwidth]{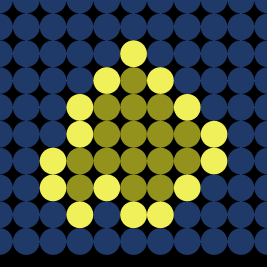}} \quad
\subfloat[][]
{\includegraphics[width=.20\columnwidth]{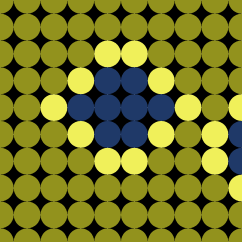}} \quad
\subfloat[][]
{\includegraphics[width=.20\columnwidth]{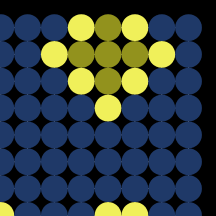}} \\
\subfloat[][]
{\includegraphics[width=.20\columnwidth]{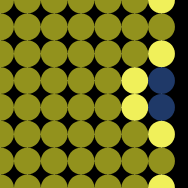}} \quad
\subfloat[][]
{\includegraphics[width=.20\columnwidth]{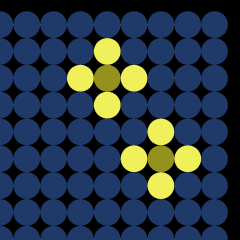}} \quad
\subfloat[][]
{\includegraphics[width=.20\columnwidth]{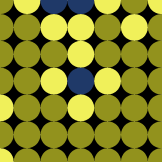}}
\caption{{\bf Regular lattice}. \Didasc Several meaningful enlargements of patterns found in Figure~\ref{snapshots}.}
\label{enlarg}
\end{figure}
%@@@@@@@@@@@@@@@@@@@@@@@@@
As one immediately sees, in all the cases presented in Figure~\ref{snapshots}, the final outcome is a situation in which the two groups are highly segregated and form stable clusters.
We could ask why, after some time, the two groups get stuck in these frozen configurations. The answer comes from a careful look at the figures. 
\\
If we look at Figure~\ref{enlarg}, in panels (a) and (b) we can see two examples of clusters (an altruistic one and a selfish one) surrounded by players of the other group, while in panels (c) and (d) we can see two examples of analogous clusters at the edge of the system.
It results that altruists in the inner parts of altruistic clusters tend to be rich, while those at the border of a selfish or an altruistic cluster tend to be poor. This comes from the fact that an altruist that is surrounded by 4 altruists has a $1/2$ probability of donating capital (that is the probability of chosing game~A), while having at least an equal probability of receiving a donation. So, an internal altruist does not lose capital as a consequence of donation mechanism and, simultaneously, he tends to gain by means of game~B.
On the other hand, an altruist at the border of a given (altruistic or selfish) cluster is surrounded by less than 4 altruists. He still has a $1/2$ probability of donating capital, but now he has a smaller probability of receiving a donation so, due to this unfavourable balance, he tends to lose capital. 
The same also happens, for analogous reasons, to altruists at the edge of the system, who therefore tend to be poor (as clearly visible in panel (d) of Figure~\ref{snapshots}). 
As a consequence of this effect, selfish players at the border of a cluster can see only poorer altruistic players and therefore never imitate them, thus remaining frozen in their selfish state. On the other hand, poor altruistic players at the border of a cluster have to face richer selfish players, but at the same time they are in contact with internal altruistic players that are richer than those selfish players, so altruistic players at the border continue to imitate their richer fellows and remain altruistic. Summarizing, such a complex dynamics fully explains the stability of the clusters of cooperators or defectors observed in panels (a--d) of Figure~\ref{enlarg}. 
Finally, in panel (e) we can see that the smallest stable pattern of altruists is a cross (in fact, an isolated altruist, having no neighbours to exchange capital with, will inevitably end to disappear), while in panel (f) we can see that even a single isolated selfish player can survive.
\\
After the regular lattice, let us take into consideration the other topologies for both AS and SAS models. For the small-world network, we performed analogous simulations in order to evaluate the final percentage of altruists as a function of the initial one.
%@@@@@@@@@@@@@@@@@@@@@@@@@
\begin{figure}[tbp]
\centering
\includegraphics[width=\columnwidth]{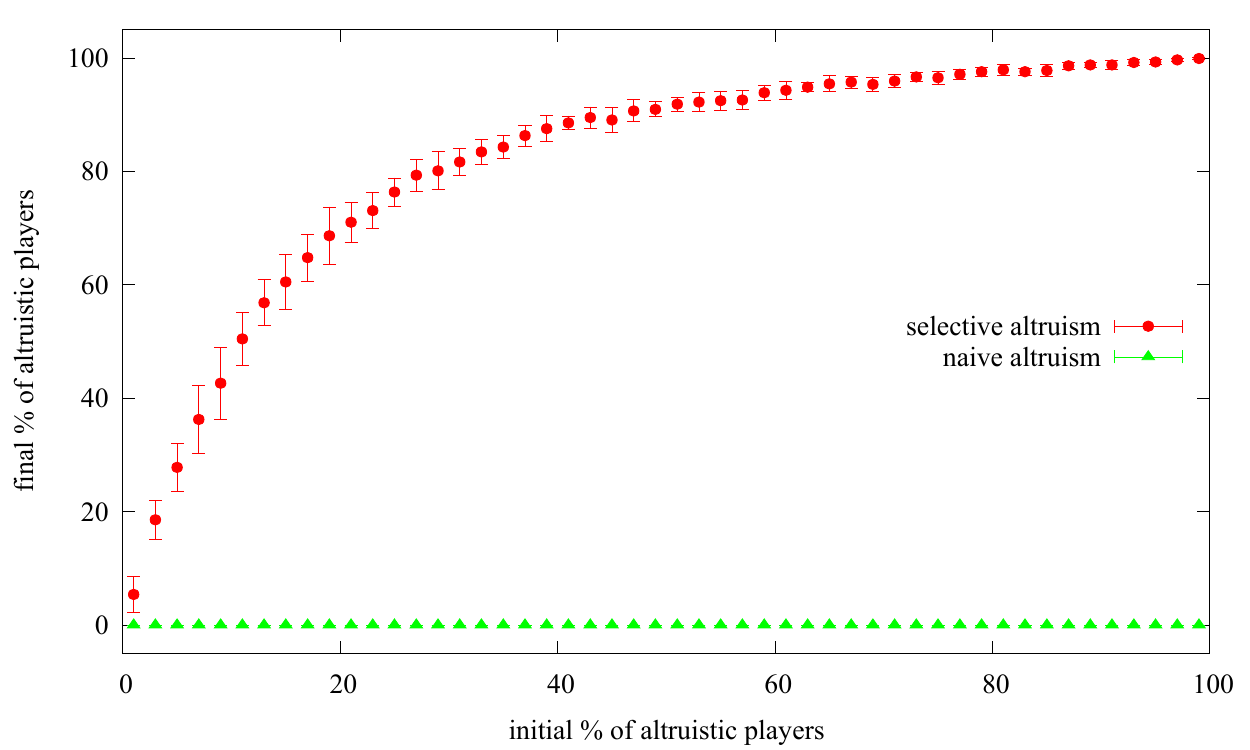}
\caption{{\bf Small-world network}. \Didasc Plot of the final percentage of altruists (taken at time 10,000) as a function of their initial percentage.
We consider AS and SAS models with imitation.
There are $N = 1936$ players, which can be selfish or altruistic, arranged in a small-world topology.
The results are averaged over 20 realizations and we report the averaged values together with bars indicating their standard deviations.}
\label{sw-network}
\end{figure}
%@@@@@@@@@@@@@@@@@@@@@@@@@
%@@@@@@@@@@@@@@@@@@@@@@@@@
\begin{figure}[tbp]
\centering
\includegraphics[width=\columnwidth]{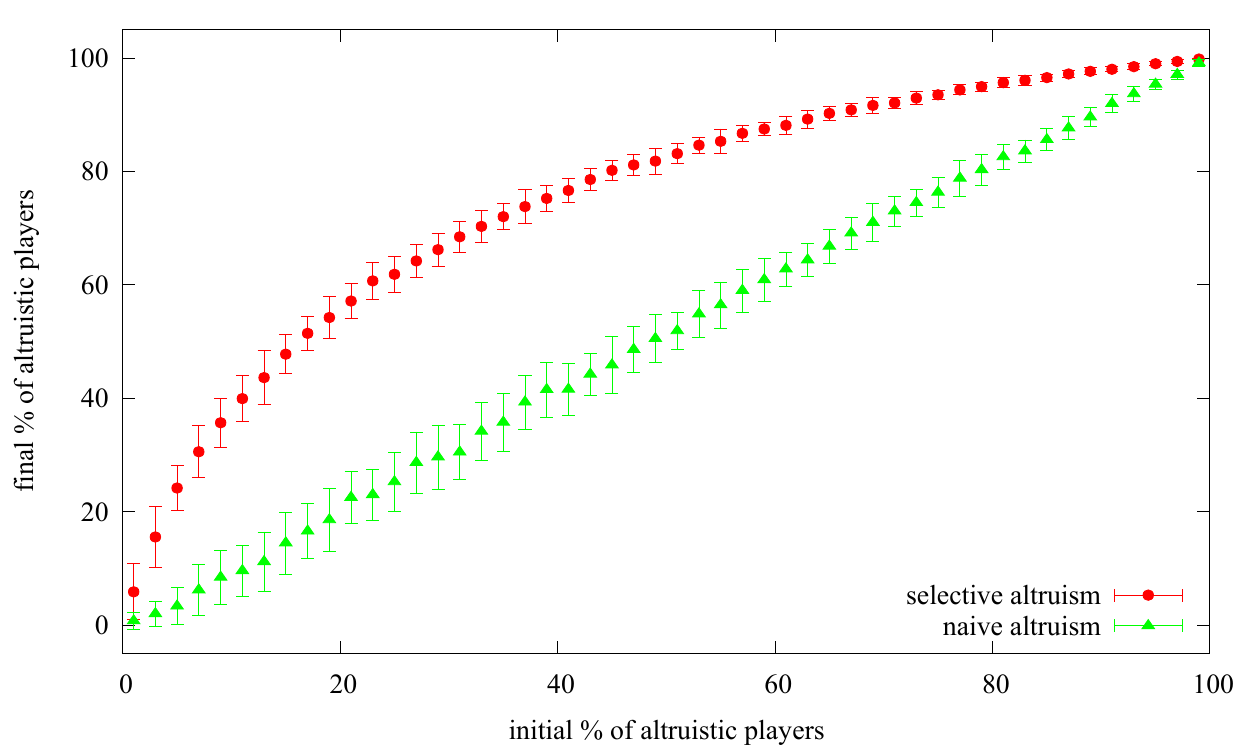}
\caption{{\bf Random network}. \Didasc Plot of the final percentage of altruists (taken at time 10,000) as a function of their initial percentage.
We consider AS and SAS models with imitation.
There are $N = 1936$ players, which can be selfish or altruistic, arranged in a random network.
The results are averaged over 20 realizations and we report the averaged values together with bars indicating their standard deviations.}
\label{rand-network}
\end{figure}
%@@@@@@@@@@@@@@@@@@@@@@@@@
The results are reported in Figure~\ref{sw-network}.
We can see that the situation is practically the same as in the regular lattice, both in the case of naive and selective altruism.
In other terms, rewiring some of the links of the regular lattice does not produce a significant change on the global diffusion of altruism. Only, the shape of stable selfish and altruistic patterns can  be deformed by the presence of long range links with respect to the regular lattice topology. 
\\
In Figure~\ref{rand-network} we can see the results for the random network topology. Selective altruism has a final diffusion that is similar to the one found for the regular lattice and for the small-world network. The only unexpected result is that, in this topology, naive altruism does not tend to disappear. On the contrary, its final percentage is nearly equal to the initial one. Examining the details of time evolution, we found that in this case, for a given starting percentage of altruists, there is an initial sudden decrease of altruistic players immediately followed by an increase of the same order. Therefore their final percentage remains almost unchanged.
%@@@@@@@@@@@@@@@@@@@@@@@@@
\begin{figure}[tbp]
\centering
\includegraphics[width=\columnwidth]{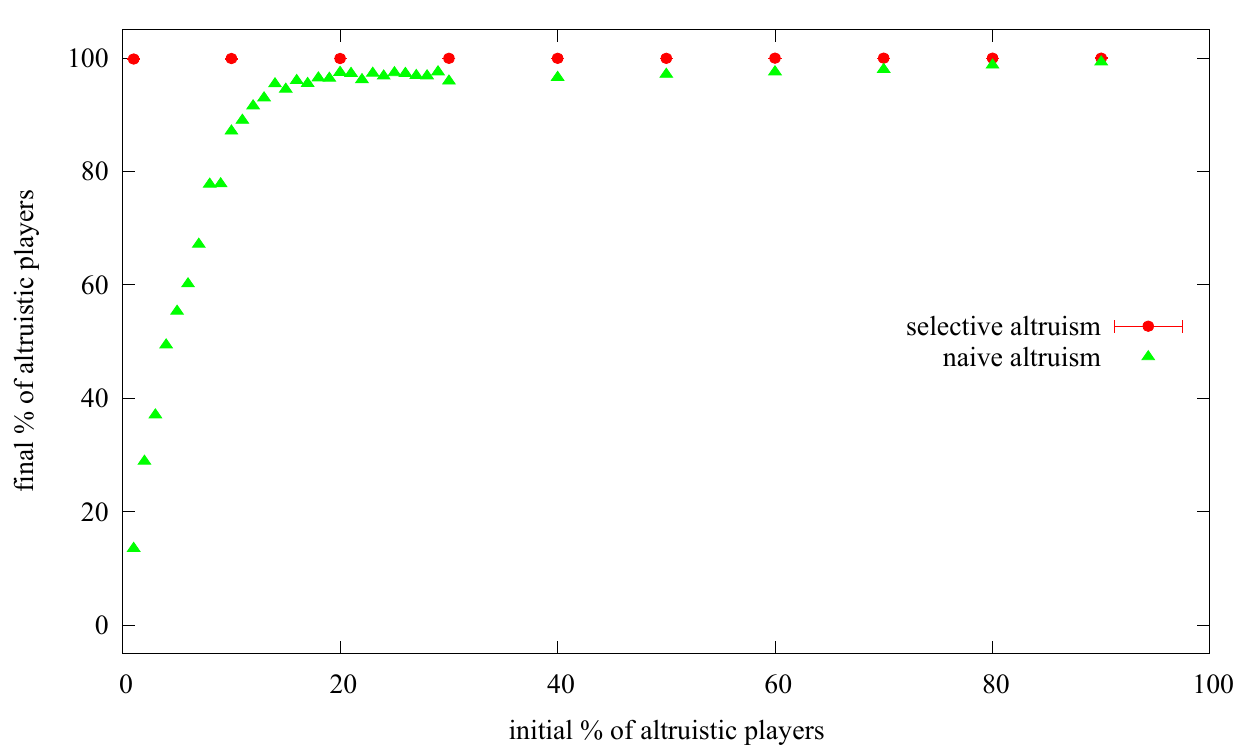}
\caption{{\bf Scale-free network}. \Didasc Plot of the final percentage of altruists (taken at time 300, since in this topology convergence is very fast) as a function of their initial percentage.
We consider AS and SAS models with imitation.
There are $N = 1936$ players, which can be selfish or altruistic, arranged in a scale-free network.
The results are averaged over 20 realizations and we report the averaged values together with bars indicating their standard deviations.}
\label{sf-network}
\end{figure}
%@@@@@@@@@@@@@@@@@@@@@@@@@
\\
The last considered topology is the scale-free network, whose results are reported in Figure~\ref{sf-network}.
Again, we see that the percentage of naive altruists does not evolve to zero. In particular, for an initial percentage above 20\% the final percentage is always near 100\% but, for smaller initial percentages, there are some realizations in which the final percentage is near 100\% and other ones in which it is near 0\%. Taking the average value over 200 different realizations, we obtain the curve that we can see in the left part of the panel. We do not report error bars because there is not a distribution of values around the most likely value, but only a binary outcome (0\% or 100\%). It seems that a value near 20\% can  play the role of a threshold for the initial percentage of altruistic players one needs to have in order to observe a complete diffusion of altruism on a scale-free network community, but this feature surely deserves a deeper analysis which will be done elsewhere. Finally, in the case of selective altruism, the number of altruists always increases to almost the entire population, thus giving rise, in average, to a constant final percentage of altruists for any initial condition.

\section*{Conclusions}
\addcontentsline{toc}{section}{Conclusions}

In this work we investigated  altruism in two modified versions of collective games exploiting Parrondo's paradox. 
First, we considered a modified version of Toral's model, the AS model, showing that a small fraction of altruistic players can give  a winning trend to the whole community of fully interacting people. However, those naive altruistic players help others at their own expense: while selfish players can enjoy a big gain of capital, altruistic players suffer a big loss. In this situation altruism is strongly discouraged. To overcome this problem, we considered, in the SAS model, a more refined and realistic  way for a player to help other players, a behaviour that we called {\it selective altruism}. The key factor  is  the process of {\it repeated encounters}. In this context, it means that two players who are connected can play together many times, so they have a reason to trust each other and help other players that help them.
Simulations showed that selective altruists create a winning trend for their category, but avoid at the same time to be exploited by selfish players. The latter are left out from this network of mutual help: they can only play a losing game and so their capital decreases. 
\\
We then introduced imitation in AS and SAS models: all the players can change their behaviour in time, imitating at each turn the most successful among them. In this case selfish players prevail against naively altruistic players, because they have much higher gains, and so all the players sooner or later become selfish. Conversely, selectively altruistic players obtain higher gains than selfish ones and are imitated by them, so all the players  become altruistic very soon.
\\
Lastly, we studied the influence of topology on the diffusion of altruism. We found that some topologies put a limit to the diffusion of selective altruism, while other topologies allow even naive altruism to diffuse among the population. In the particular case of a regular lattice we also discovered interesting emerging patterns of altruistic and selfish behaviours in the final stationary configurations of the system. The latter  have  a simple explanation in terms of exchanges of capital among players.    
\\
In conclusion, by means of extensive numerical simulations with collective kinds of  Parrondo's games, we showed how, as often observed in real situations,  a community  of cooperators can be  favoured with  respect to a community of defectors, because cooperation can  become a  winning strategy for the group.  With a mechanism of reputation, altruistic players can be aware of what players are worthy of trust and so they can overcome the negative effects of being naively altruistic. In this situation, a process similar to natural selection acts on the population and produces a socio-economic evolution \cite{helbing3}: altruists become richer, their behaviour is imitated and spreads among the population, extending in this way its positive effects to all the players.

\section*{Acknowledgments}
\addcontentsline{toc}{section}{Acknowledgments}

We thank Dirk Helbing for several comments  and suggestions.

\end{document}